\documentclass[12pt]{article}
\usepackage{graphicx,epsfig}
\usepackage{amsmath,textcomp}
\usepackage{latexsym}
\usepackage{amstext}

\usepackage{cite}

\usepackage{ulem}
\usepackage{amssymb}
\usepackage{array}
\usepackage{multirow}
\usepackage{tikz}

\newcommand{\mb}[1]{ \mbox{\boldmath$#1$} }
\newcommand{\ds}{\displaystyle}
\newcommand{\beq}{\begin{eqnarray}}
\newcommand{\eeq}{\end{eqnarray}}
\newcommand{\beqq}{\begin{eqnarray*}}
\newcommand{\eeqq}{\end{eqnarray*}}
\newcommand{\p}{\partial}

\newcommand{\eps}{\varepsilon}
\newcommand{\x}{\mbox{\boldmath$x$}}

\newcommand{\A}{\mbox{\boldmath$A$}}
\newcommand{\y}{\mbox{\boldmath$y$}}
\newcommand{\z}{\mbox{\boldmath$z$}}
\newcommand{\n}{\mbox{\boldmath$n$}}

\font\bb=msbm10 at 12pt

\def\rR{\hbox{\bb R}}

\begin{document}

\title{Asymptotics of extreme statistics of escape time in 1,2 and 3-dimensional diffusions}

\author{K. Basnayake\footnote{Applied Mathematics \& Computational Biology, Ecole Normale Sup\'erieure, 46 rue d'Ulm 75005 Paris, France.} , C. Guerrier\footnote{The University of British Columbia Mathematics Department, Vancouver, B.C. Canada V6T 1Z2} , Z. Schuss\footnote{Department of Applied Mathematics, Tel-Aviv University, Tel-Aviv 69978, Israel.} , D. Holcman\footnotemark[1]}
\date{}
\maketitle
\begin{abstract}
The first of $N$ identical independently distributed (i.i.d.) Brownian trajectories that arrives to a small target, sets the time scale of activation, which in general is much faster than the arrival to the target of only a single trajectory. Analytical asymptotic expressions for the minimal time is notoriously difficult to compute in general geometries. We derive here asymptotic laws for the probability density function of the first and second arrival times of a large number of i.i.d. Brownian trajectories to a small target in 1,2, and 3 dimensions and study their range of validity by stochastic simulations. The results are applied to activation of biochemical pathways in cellular transduction.
\end{abstract}

\section{Introduction}
Fast activation of biochemical pathways in cell biology is often initiated by the first arrival
of a particle to a small target. This is the case of calcium activation in synapses of neuronal
cells \cite{Guerrier,Volfovsky,BJ2005}, fast photoresponse in rods, cones and fly photoreceptors
\cite{Minke,Burns,Reingruber}, and many more. However, the time scale underlying these fast
activations is not very well understood. We propose here that these mechanisms are initiated by
the first arrival of one or more of the many identical independently distributed (i.i.d.)
Brownian particles to small receptors (such as the influx of many calcium ions inside a synapse
to receptors).

In general, one or several particles are required to initiate a cascade of chemical reactions,
such as the opening of a protein channel \cite{Hille} of a cellular membrane, which amplifies the
inflow of ions to an avalanche of thousand or more ions, resulting from binding of
couple of ions. These statistic of the minimal arrival times are referred to in the statistical
physics literature as extreme statistics \cite{extreme1}. Despite great efforts
\cite{extreme1,extreme2,extreme3,extreme4,extreme5,extreme6,extreme7,extreme8,extreme9}, there are no explicit
expressions for the probability distributions of arrival times of the first trajectory, the second, and so on. Only general formulas are given, and they account for neither specific geometrical constraints of the bounding domains, where particles evolve, nor for the small targets that bind these particles.

The main example to keep in mind is the statistics of the escape time
through a small window of the first of $N$ particles. Asymptotic expressions for the mean escape
time of a single Brownian path, the so called narrow escape time, computed in the
narrow escape theory \cite{holcmanschuss2015,SIREV},  depends on global and local
geometric properties of the bounding domain and its boundary, such as the surface area (in 2
dimensions or volume (in 3 dimensions), and the local geometry near the absorbing
window (mean curvature of the boundary at the small window, the window's shape, and relative
size).  The number and distribution of absorbing windows can influence
drastically the narrow escape time. As shown below, the escape of the fastest
particles selects trajectories that are very different from the typical ones, which
determine the mean narrow escape time (NET).

Moreover, the asymptotics of the expected first arrival time are not the same as of the mean first passage
time (MFPT) of a single Brownian path to a small window. The analysis relies on the time-dependent solution of the Fokker-Planck equation and the short-time asymptotics of the survival probability.
Previous studies of the short-time asymptotics of the diffusion equation concern the asymptotics
of the trace of the heat kernel analysis \cite{CollindeVerdiere,Spivak}. Here, an estimate is
needed of the survival probability, which requires different analysis. Our method is based on the
construction of the asymptotics from Green's function of the Helmholtz equation.

The main results are explicit expressions for the statistics of the first arrival
time see attempt in \cite{extreme7}) in 1,2, and 3 dimensions and a formula for the expected
shortest exit time from a neuronal spine with and without returns. The manuscript is organized as follows. First we present the general framework for the computation of the pdf of the first arrival and conditional second arrival, given that the first one has already arrived, in a population of $N$ Brownian particles in a ray and in an interval. We then study the difference between Poissonian and diffusion escape time approximations and in particular, we consider the case of a bulk domain with a window connected to a narrow
cylinder (dendritic spine shape \cite{holcmanschuss2015,SIREV}). We then compute the pdf of the
extreme escape time in dimensions 2 and 3 through small windows. Finally, we discuss
applications to activation in cellular biology.
\section{The pdf of the first escape time}
The narrow escape problem (NEP) for the shortest arrival time of $N$  non-interacting
i.i.d. Brownian trajectories (ions) in a bounded domain $\Omega$ to a binding site is defined as follows. Denote by $t_i$ the arrival times and by $\tau^{1}$ the shortest one,
\beq
\tau^{1}=\min (t_1,\ldots,t_N),
\eeq
where $t_i$ are the i.i.d. arrival times of the $N$ ions in the medium. The NEP is to find the PDF and the MFPT of $\tau^{1}$. The complementary PDF of $\tau^{1}$ is given by
\beq
\Pr\{\tau^{1}>t \} = {\Pr}^N\{t_{1}>t \},
\eeq
where $\Pr\{t_{1}>t \}$ is the survival probability of a single particle prior to binding at the
target. This probability can be computed from the following boundary value problem. Assuming that
the boundary $\p\Omega$ contains $N_R$ binding sites $\p\Omega_i\subset\p \Omega\
(\p\Omega_a=\bigcup\limits_{i=1}^{N_R}\p\Omega_i,\ \p\Omega_r=\p\Omega-\p\Omega_a)$, the pdf of a
Brownian trajectory is the solution of the  initial boundary value problem (IBVP)
\begin{align}\label{IBVP}
\frac{\p p(\x,t)}{\p t} =&D \Delta p(\x,t)\quad\hbox{\rm for } \x \in \Omega,\ t>0\\
p(\x,0)=&p_0(\x)\quad \hbox{\rm for } \x \in\Omega\nonumber\\
\frac{\p p(\x,t)}{\p \n} =&0\quad \hbox{\rm for } \x \in\p\Omega_r\nonumber\\
p(\x,t)=&0\quad \hbox{\rm for } \x \in \p\Omega_a.\nonumber
\end{align}
The survival probability is
\beq \label{surv}
\Pr\{t_{1}>t \} =\int\limits_{\Omega} p(\x,t)\,d\x,
\eeq
so that
\beq
\Pr\{\tau^{1}=t \} = \frac{d}{dt}\Pr\{\tau^{1}<t \}=N(\Pr\{t_{1}>t \})^{N-1}\Pr\limits\{t_{1}=t
\},
\eeq
where
\beq
\Pr\{t_{1}=t \}= \oint\limits_{\p\Omega_a} \frac{\p p(\x,t)}{\p \n}\, dS_{\x}.
\eeq
\beq
\Pr\{t_{1}=t \}= N_R \oint\limits_{\p \Omega_1} \frac{\p p(\x,t)}{\p \n}\,dS_{\x}.
\eeq
Putting all the above together results in the pdf
\beq \label{arrv1}
\Pr\{\tau^{1}=t \} =N N_R  \left[\int\limits_{\Omega} p(\x,t)d\x \right]^{N-1}\oint\limits_{\p \Omega_1} \frac{\p p(\x,t)}{\p \n} dS_{\x}.
\eeq
The first arrival time is computed from the survival probability of a particle and the flux through the
target. Obtaining an explicit or asymptotic expression is not possible in general.
\subsection{The pdf of the first arrival time in an interval}
\begin{figure}[http!]
\centering
\includegraphics[width=.99\textwidth]{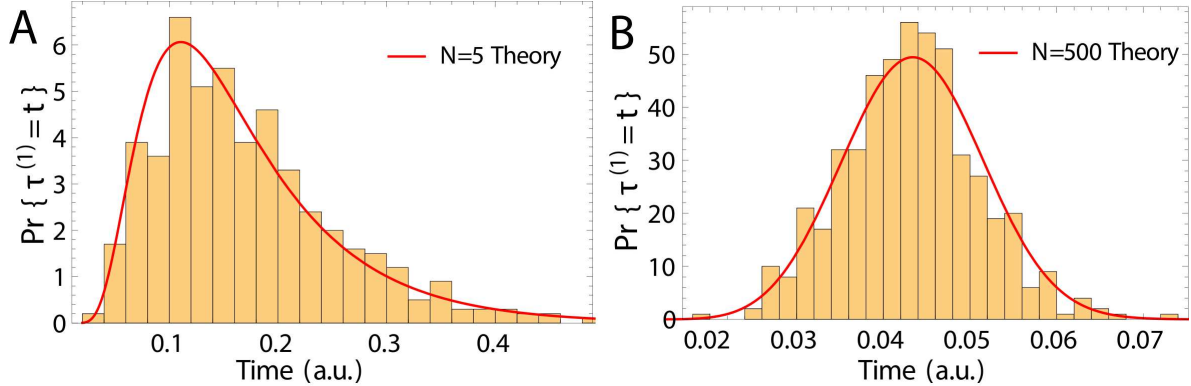}
\caption[caption]{\small Histograms of the arrival times to the boundary of the fastest particle, obtained from Brownian simulations with Euler's scheme. The number of Brownian particles  is $N=5$ in \textbf{A} and  $N=500$ in \textbf{B}. The analytical solution (red curves) is obtained by setting $n_0=100$ in \eqref{eq:approx}.}\label{fig:0}
\end{figure}

To obtain an analytic expression for the pdf of the first arrival time \eqref{arrv1} of a particle inside a narrow neck that could represent the  dendritic spine neck, we model the narrow spine neck as a segment of length $L$, with a reflecting boundary at $x=0$ and absorbing boundary at $x=L$. Then the diffusion boundary value problem \eqref{IBVP} becomes
\begin{align}
\ds{\frac{\p p}{\p t}}=& \ds{ D \frac{\p^2 p}{\p x^2}}\quad\mbox{  for}\ 0<x<L,\ t>0\label{pde1}
 \\
p(x,0)=& \delta(x)\quad\mbox{  for}\ 0<x<L\\
p(L,t)=& \frac{\p p(0,t)}{\p x} =0\quad \hbox{ for } t>0,
\end{align}
where the initial condition corresponds to a particle initially at the origin. The general solution is given by the eigenfunction expansion
\beq \label{sol}
p(x,t)= 2 \sum^{\infty}_{n=0} e^{-D \lambda^2_nt} \cos\lambda_n x,
\eeq
where the eigenvalues are
\beq
\lambda_n=\frac{\pi}{L}\left(n+\frac{1}{2}\right).
\eeq
The survival probability \eqref{surv} of a particle is thus given by
\beq\label{eq:prtau1grthan}
\Pr\{t_{1}>t \}=\int\limits_{0}^{L} p(x,t)dx =2 \sum^{\infty}_{n=0} \frac{(-1)^{n}}{\lambda_n}e^{-D\lambda^2_nt}.
\eeq
The pdf of the arrival time to $L$ of a single Brownian trajectory is the probability efflux at the absorbing boundary $\p\Omega_a$, given by
\beq
-\oint\limits_{\Omega_a} \frac{\p p(\x,t)}{\p \n}\, dS_{\x}=-\frac{\p p(L,t)}{\p x}=
2 \sum^{\infty}_{n=0} {(-1)^{n}}{\lambda_n}e^{-D \lambda^2_nt}.
\eeq
Therefore, the pdf of the first arrival time in an ensemble of $N$ particles to one of $N_R$ independent absorbers is given by
\beq\label{eq:tau1}
\Pr\{\tau^{(1)}=t\}=2NN_R\left(2 \sum^{\infty}_{n=0} \frac{(-1)^{n}}{\lambda_n}e^{-D \lambda^2_nt}\right)^{N-1} \sum^{\infty}_{n=0} {(-1)^{n}}{\lambda_n}e^{-D \lambda^2_nt}.
\eeq
For numerical purposes, we approximate \eqref{eq:tau1} by the sum of $n_0$ terms,
\beq
\label{eq:approx}
\Pr\{\tau^{(1)}=t\}\approx f_{n_0}(t)= NN_R\left(\sum^{n_0}_{n=0} \frac{(-1)^{n}}{\lambda_n}e^{-D \lambda^2_nt}\right)^{N-1} \sum^{n_0}_{n=0} {(-1)^{n}}{\lambda_n}e^{-D \lambda^2_nt}.
\eeq
Figs.\ref{fig:0}A-B show the pdf of the first arrival time for $N=5$ and $N=500$
Brownian particles with diffusion coefficient $D=1$, which start at $x=0$ at time 0
and exit the interval at $x=1$. These figures confirm the validity of the analytical
approximation \eqref{eq:tau1} with only $n_0=100$ terms in the lowly converging alternating
series.

\section{Statistics of the arrival time of the second particle}\label{s:stat}
Next, we turn to the computation of the conditional pdf of the arrival time
$\tau^{(2)}$ of the second particle, which is that of the minimum of the shortest arrival time in
the ensemble of $N-1$ trajectories after the first one has arrived, conditioned on their locations at time $\tau^{(1)}$. The time $\tau^{(1)}+\tau^{(2)}$ is that of arrival of the first two particles at reach the target.  The conditional distribution of the arrival time $\tau^{(2)}$ of the second particle, given the positions of the $N-1$ particles at time $\tau^{(1)},$ can be computed from their joint probability distribution at positions $(x_1,\ldots,x_{N-1})$ and the first particle has already arrived at time $\tau^{(1)}=s$,
\beq \label{Prtau2}
&&\Pr\{\tau^{(2)}=t\}\\
&=&\int_{0}^{t} \int_{\Omega}..\int_{\Omega} \Pr\{\tau^{(2)}=t,  \tau^{1}=s, \x_1(s)=x_1,\ldots,\x_{N-1}(s)=x_{N-1} \}\,dx_{1}\cdots\,dx_{N}\,ds \nonumber
\label{tau2a}
\eeq
and
\beqq
&&\Pr\{\tau^{(2)}=t,  \tau^{1}=s, x_1(s)=x_1,\ldots,x_{N-1}(s)=x_{N-1} \}\\
&=&\Pr\{\tau^{(2)}=t\,|\,\tau^{1}=s, x_1(s)=x_1,\ldots, x_{N-1}(s)=x_{N-1} \}  \nonumber\\
&& \times\Pr\{\tau^{1}=s\} \Pr\{ x_1(s)=x_1,\ldots,x_{N-1}(s)=x_{N-1} \}. \nonumber
\eeqq
Because all particles are independent,
\beqq
\Pr\{ \x_1(s)=x_1,\ldots,\x_{N-1}(s)=x_{N-1} \}=\prod_{i=1}^{N-1} \Pr\{ \x_i(s)=x_i\}
\eeqq
so that
\beq \label{fulltau2}
\Pr\{\tau^{(2)}=t\}=\int\limits_{0}^{t}  \Pr\{\tau^{(2)}=t\,|\,  \tau^{1}=s\} \left(\int\limits_{0}^{L} \Pr\{\x_1(s)=x_1\}dx_1\right)^N \Pr\{\tau^{1}=s \}ds.
\eeq

\subsection{The Poissonian-like approximation}
The pdf \eqref{fulltau2} can be evaluated under some additional assumptions. For
example, if the Brownian trajectories escape from a deep potential well,  the escape process is
well approximated by a Poisson process with rate equal the reciprocal of the mean
escape time from the well \cite{DSP}. Also, when Brownian particles escape a domain $\Omega=B\cup
C$, which consists of a bulk $B$ and a narrow cylindrical neck $C$, the escape process from
$\Omega$ can be approximated by a Poisson process, according to the
narrow scape theory \cite{holcmanschuss2015}. Here the motion in the narrow cylinder $C$ is approximated by one-dimensional Brownian motion in an interval of length $L$.

Consequently, under the Poisson approximation, the arrival of the first particle is much faster than the escape of the second one from the bulk compartment $B$, thus we can use the approximation that all particles are still in the bulk $B$ after the arrival of the first one. The bulk is represented by the position $x=0$ in an approximate one-dimensional model. This assumption simplifies \eqref{fulltau2} to
\beq \label{approx}
\left(\int\limits_{0}^{L} \Pr\{\x_1(s)=x_1\}dx_1\right)^N \approx 1,
\eeq
so that
\beq
\Pr\{\tau^{(2)}=t\}=\int\limits_{0}^{t} \Pr\{\tau^{(2)}=t|\tau^{1}=s\}\Pr\{\tau^{1}=s\}ds.
\eeq
The Markovian property of the Poisson process gives
\beq \label{noapprox}
\Pr\{\tau^{(2)}=t|\tau^{1}=s\}=\Pr\{\tau^{(2)}=t-s,\}
\eeq
so that $\tau^{(2)}$ has the same pdf as $\tau^{1}$ with  $N-1$ particles, which we approximate to be the same for large $N$, that is,
\beq\label{time2}
\Pr\{\tau^{(2)}=t\}=\int\limits_{0}^{t} f(t-s)f(s)\,ds,
\eeq
where (recall \eqref{eq:tau1})
\beq\label{tau1expr}
f(s)=\Pr\{\tau^{1}=s\}=N N_R g(t)^N h(t).
\eeq
In one dimension, $g(t)=\sum^{N_t}_{n=0} \frac{(-1)^{n}}{\lambda_n}e^{-D \lambda^2_nt}$ and  $h(t)=\sum^{N_t}_{n=0} {(-1)^{n}}{\lambda_n}e^{-D \lambda^2_nt}$. It follows that
\beq \label{secondaversuss}
\Pr\{\tau^{(2)}=t \}\approx N^2N_R^2\int\limits_{0}^{t} g(s)^{N-1} h(s) g(t-s)^{N-1} h(t-s)ds.
\eeq
\subsection{$\Pr\{\tau^{(2)} \}$ of $N$ Brownian i.i.d. trajectories in a segment}
\begin{figure}[http!]
\centering
\includegraphics[width=.69\textwidth]{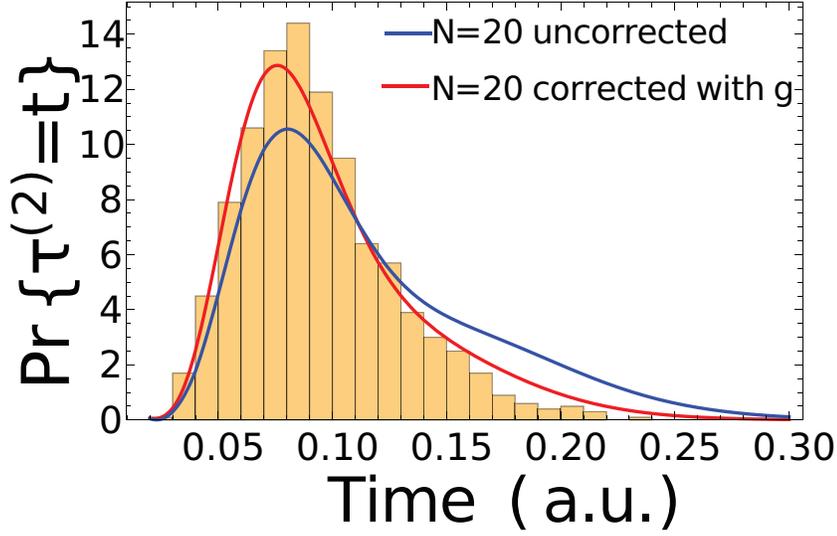}
\caption[caption]{\small Histogram of the arrival time of the second fastest particle, obtained from Brownian simulations with Euler's scheme. The fastest is computed for $N=20$ in \textbf{B.} The analytical solution with no correction is given by  \eqref{noapprox} (blue) and compared to \eqref{approx2} with the correction (red). There are $n_0=6$ terms in the series \eqref{eq:approx}.}\label{fig:t2}
\end{figure}

{As in the first paragraph of section \ref{s:stat}, equations \eqref{Prtau2} and \eqref{fulltau2} are valid with $\Omega$  replaced by
the segment $[0,L]$. That is,
\beqq
&&\Pr\{\tau^{(2)}=t\}\\
&=&\int\limits_{0}^{t} \int\limits_{0}^{L}\cdots\int\limits_{0}^{L} \Pr\{\tau^{(2)}=t,  \tau^{1}=s, x_2(s)=x_2,\ldots,x_{N}(s)=x_{N} \}\,dx_{2}\cdots\,dx_{N}\,ds.
\eeqq
and
\beqq
&&\Pr\{\tau^{(2)}=t,  \tau^{1}=s, x_2(s)=x_1,\ldots,x_{N-1}(s)=x_{N} \}\\
&= &\Pr\{\tau^{(2)}=t | \tau^{1}=s, x_2(s)=x_1,\ldots,x_{N-1}(s)=x_{N} \}  \nonumber\\
&& \times\Pr\{\tau^{1}=s\} \Pr\{ x_2(s)=x_1,\ldots,x_{N-1}(s)=x_{N} \}. \nonumber
\eeqq
Because all particles are independent,
\beq
\Pr\{ x_2(s)=x_1,\ldots,x_{N}(s)=x_{N-1} \}=\prod_{i=1}^{N-1} \Pr\{ x_{i+1}(s)=x_i\},
\eeq
hence,
\beq \label{tau2}
\Pr\{\tau^{(2)}=t\}=\int\limits_{0}^{t}  \Pr\{\tau^{(2)}=t|  \tau^{1}=s\} \left(\int\limits_{0}^{L} \Pr\{x_1(s)=x_1\}dx_1\right)^{N -1} \Pr\{\tau^{1}=s \}\,ds.
\eeq
}
To compute the survival probability
\beq \label{survival}
S(s)=\int\limits_{0}^{L} \Pr\{x_1(s)=x_1\}\,dx_1,
\eeq
we use the short-time asymptotics of the one-dimensional diffusion equation. {Modifying
 equation \eqref{pde1} for short-time diffusion of a particle} starting at $0$, we get
\begin{align}\label{pde11}
\frac{\p p(x,t)}{\p t} =&D \frac{\p^2 p(x,t)}{\p x^2}\quad\mbox{for}\ x>0,\ t>0 \nonumber\\
p(x,0)=&\delta(x)\quad\mbox{for}\ x>0,\quad p(L,t)=0\quad\mbox{for}\ t>0.  
\end{align}
{The short-time diffusion is well approximated by the fundamental solution (except at the boundary, where the error is exponentially small in $1/t$)
\beq
p(x,t) =\frac{1+o(t)}{\sqrt{4D \pi t}}\exp\left\{ - \frac{x^2}{4Dt}\right\}.
\eeq
Thus the survival probability at short time $t$ is
\beq \label{survival2}
S(t)= \int\limits_{0}^{L} \frac{1+o(t)}{\sqrt{4D \pi t}}\exp\left\{ - \frac{x^2}{4Dt}\right\}\,dx.
\eeq
The short-time asymptotic expansion \eqref{asymp} (see below) and the change of variable $x=u\sqrt{4Dt}$ in the integral \eqref{survival2}, give
\beq
S(t)&=&1- \frac{1}{\sqrt{\pi }}\int\limits^{\infty}_{L/\sqrt{4Dt}} \left[\exp\left\{ - u^2\right\}\right] du\\
&\approx &1- \sqrt{4Dt}\frac{\exp\left\{ - (L/\sqrt{4Dt})^2\right\}}{\sqrt{\pi } L} \left(1-2\frac{Dt}{L^2}+ O\left(\frac{t^2}{L^4}\right)\right).
\eeq
It follows from  \eqref{tau2} that the pdf of the second arrival time is
\beq \label{approx2}
&&\Pr\{\tau^{(2)}=t\}\\
&=&[1+o(1)]\int\limits_0^t  \Pr\{\tau^{1}=s\} \Pr\{\tau^{1}=t-s\} \left(1- \sqrt{4Ds}\frac{\exp\left\{ - (\frac{L}{\sqrt{4Ds}})^2\right\}}{\sqrt{\pi } L} \right)^{N-1}\,ds.\nonumber
\eeq
Figure \ref{fig:t2} compares results of the stochastic simulations with the analytical formula
\eqref{secondaversuss} for the second fastest arrival time $\tau^{(2)}$ to the boundary 1 of the
interval $[0,1]$ among 20 particles. We use the analytical formula \eqref{secondaversuss}  (no
correction) and \eqref{approx2}, which contains the shift correction due to the distribution of the particles in the interval at time $\tau^{(1)}$, when the first particle arrives at $x=L$ been absorbed.}

This figure shows how the corrected formula gives a better agreement with the Brownian
simulations, thus proving that the distribution of the Brownian particles inside the interval
contributes to the decrease of the arrival time of the second particle. The fit to simulation
data is based on the eigenfunction expansion
\beq \label{approx22}
\Pr\{\tau^{(2)}=t\}=[1+o(1)]\int\limits_{0}^{t}  \Pr\{\tau^{1}=s\} \Pr\{\tau^{1}=t-s\} \left(2 \sum^{\infty}_{n=0} \frac{(-1)^{n}}{\lambda_n}e^{-D\lambda^2_nt} \right)^{N-1}ds,
\eeq
which is equivalent to \eqref{approx2}. Formula \eqref{tau1expr} is used for $\Pr\{\tau^{1}=s\}$.  Note that the alternating series contains an even number of terms.

To conclude, the internal distribution of particle is given by $$\left(1-
\sqrt{4Ds}\frac{\exp\left\{ - (\frac{L}{\sqrt{4Ds}})^2\right\}}{\sqrt{\pi } L}
\right)^{N-1}$$ and causes the faster arrival of the second particle relative to the first one. This example shows the deviation from the purely Poissonian approximation.
\section{Asymptotics of the expected shortest time $\bar{\tau}^{1}$}
The MFPT of the first among $N$ i.i.d. Brownian {paths} is given by
\beq\label{mfptmin}
\bar{\tau}^{1}=\int\limits\limits_0 ^{\infty}\Pr\{\tau^{1}>t\} dt = \int\limits_0 ^{\infty} \left[ \Pr\{t_{1}>t\} \right]^N dt,
\eeq
where $t_{1}$ is the arrival time of a single Brownian {path}. Writing the last integral in \eqref{mfptmin} as
\beq
\bar{\tau}^{1} =\int\limits_0 ^{\infty}e^{N \ln g(t)}dt,
\label{tau1int}
\eeq
it can be expanded for $N\gg1$ by  Laplace's method. Here
\beq
g(t)=\sum^{\infty}_{n=0} \frac{(-1)^{n}}{\lambda_n}e^{-D \lambda^2_nt}
\eeq
(see  \eqref{eq:prtau1grthan}).
\subsection{Escape from a ray}\label{rayescape}
Consider the case $L=\infty$ and the IBVP
\beq
\frac{\p p(x,t)}{\p t}& =&D \frac{\p^2 p(x,t)}{\p x^2}\quad\mbox{ for } x>0,\ t>0 \nonumber\\
p(x,0)&=&\delta(x-a)\quad\mbox{ for }\ x>0,\quad p(0,t)=0\quad\mbox{ for } t>0,
\eeq
whose solution is
\beq
p(x,t) =\frac{1}{\sqrt{4D \pi t}}\left[\exp\left\{ - \frac{(x-a)^2}{4Dt}\right\}-
\exp\left\{ - \frac{(x+a)^2}{4Dt}\right\}\right].
\eeq
The survival probability with $D=1$ is
\beq\label{eq:prtau1grthan3}
\Pr\{t_{1}>t \}=\int\limits\limits_{0}^{\infty} p(x,t)\,dx=1-\frac{2}{\sqrt{\pi}} \int\limits\limits_{a/\sqrt{4t}}^{\infty}e^{-u^2}\,du.
\eeq
To compute the MFPT in \eqref{mfptmin}, we use the expansion of the complementary error function
\beq \label{asymp}
\frac{2}{\sqrt{\pi}} \int\limits\limits_{x}^{\infty}e^{-u^2}\,du =\frac{e^{-x^2}}{x\sqrt{\pi}}\left(1-\frac{1}{2x^2}+O(x^{-4})\right)\quad\mbox{for}\ x\gg1,
\eeq
which gives
\beq
I_N\equiv \int\limits\limits_0 ^{\infty} \left[ \Pr\{t_{1}>t\} \right]^N dt \approx \int\limits\limits_0 ^{\infty} \exp\left\{ N\ln\left(1-\frac{e^{-(a/\sqrt{4t})^2}}{(a/\sqrt{4t})\sqrt{\pi}}\right)\right\}\, dt ,
\eeq
and with the approximation
\beq \label{approxI1}
I_N\approx \int\limits\limits_0^{\infty} \exp \left\{ -N\frac{\sqrt{4t}e^{-\frac{a^2}{4t}}}{a\sqrt{\pi}} \right\}\, dt=
\frac{a^2}{4}\int\limits\limits_0^{\infty} \exp \left\{ -N\frac{\sqrt{u}e^{-\frac{1}{u}}}{\sqrt{\pi}} \right\}du.
\eeq
To evaluate the integral \eqref{approxI1}, we make the monotone change of variable
\beq
w=w(t)=\sqrt{t}e^{-1/t}, \quad
w'(t)=\sqrt{t}e^{-\frac{1}{t}}\left(\frac{1}{2t} +\frac{1}{t^2}\right).
\eeq
Note that for small $t$,
\beq
w'(t)\approx w \frac{1}{t^2}
\eeq
and   $\ln w \approx -1/t$. Thus,
\beq
w'(t)\approx w (\ln w)^2.
\eeq
Breaking with $N'=\frac{N}{\sqrt{ \pi}}$
\beqq
I_N &\approx& \frac{a^2}{4}\int\limits_0^{\infty} \exp \{ -N' w\} \frac{1}{\frac{dw}{dt}}\,dw\\
  &\approx& \frac{a^2}{4} \left( \int\limits\limits_0^{\delta} \exp \{ -N' w\} \frac{a^2}{w (\ln(w))^2}\, dw +\int\limits\limits_\delta^{\infty} \exp \{ -N' w\} \frac1{\frac{dw}{dt}}\, dw \right)
\eeqq
for some $0<\delta<1$, the second integral turns out to be exponentially small
in $N$ and is thus negligible relative to the first one. Integrating by parts,
\beqq
I_N &\approx& \frac{a^2}{4} \int\limits_0^{\delta} \exp \{ -N' w\} \frac{1}{w(\ln(w))^2} dw \\
 &\approx& O(\exp(-aN))+\frac{a^2}{4}N'\int\limits_0^{\delta} \exp \{ -N' w\} \frac{1}{ \ln |w|\,} dw
\eeqq
and changing the variable to $u=N'w$, {we obtain}
\beqq
N'\int\limits\limits_0^{\delta} \exp \{ -N' w\} \frac{a^2}{4 \ln|w|} \,dw =\int\limits_0^{N'\delta} \frac{a^2 \exp \{ -u \}}{4 |\ln u/N'|} \,du.
\eeqq
Expanding
\beqq
\frac{1}{|\ln u/N'|}=\frac{1}{\ln N'}\left(1+\frac{|\ln u|}{\ln N} +O\left(\frac{|\ln u|}{\ln N'}\right)^2\right)
\eeqq
for $u>\eps>0$, we obtain,
\beqq
N'\int\limits\limits_0^{\delta} \exp \{ -N' w\} \frac{a^2}{4 \ln w} dw \approx\int\limits\limits_0^{N'\delta} \exp \{ -u \}
\frac{a^2}{4
|\ln N'|}\left(1+\frac{|\ln u|}{\ln N'}\right) du.
\eeqq
Thus, breaking the integral into two parts, from $[0,\eps]$ (which is negligible) and $[\eps,\infty[$, we get
\beq
\bar{\tau}^{1} \approx \frac{a^2}{4D\ln \frac{N}{\sqrt{\pi}}}\quad\mbox{for}\ N\gg1.\label{taubarN}
\eeq
\subsection{Escape for the second fastest from half a line}
Equation \eqref{time2} and the approximation \eqref{approx} give
\beq\label{time2p}
\Pr\{\tau^{(2)}=t\}=\int\limits_{0}^{t} f(t-s)f(s)\,ds,
\eeq
where
\beq
f(s)=\Pr\{\tau^{1}=s\}=N N_R g(t)^N h(t).
\eeq
According to \eqref{eq:prtau1grthan3},
\beq\label{eq:prtau1grthan4}
g(t)=\Pr\{t_{1}>t \}=\int\limits\limits_{0}^{\infty} p(x,t)\,dx=1-\frac{2}{\sqrt{\pi}} \int\limits\limits_{a/\sqrt{4t}}^{\infty}e^{-u^2}\,du \approx
1- \frac{e^{-(a/\sqrt{4t})^2}}{(a/\sqrt{4t})\sqrt{\pi}}
\eeq
and
\beq\label{eq:prtau1grthan4b}
h(t)=-\frac{d\Pr\{t_{1}>t \}}{dt}=\frac{2}{\sqrt{\pi}} \frac{a^2}{4\sqrt{t^3}} e^{-(a/\sqrt{4t})^2}.
\eeq
The MFPT of the second arrival is obtained from \eqref{time2p} as
\beq\label{time2m}
\bar{\tau}^{(2)}&=&\int\limits_{0}^{\infty} t \Pr\{\tau^{(2)}=t\}dt =\int\limits_{0}^{\infty} t \int\limits_{0}^{t} f(t-s)f(s)ds dt\\
&=&2 \int\limits_{0}^{\infty} s f(s)ds \left(\int\limits_{0}^{\infty} f(t)dt\right) =2 \bar{\tau}^{(1)}\left(\int\limits_{0}^{\infty} f(t)dt\right)=2 \bar{\tau}^{(1)}.
\eeq
\section{Escape from an interval $[0,a]$}\label{s:rayescape}
We follow the steps of the previous section, where Green's function for the homogenous IBVP
 is now given by the infinite sum
\beq
p(x,t\,|\,y) =\frac{1}{\sqrt{ 4 D \pi t}}\sum\limits_{n=-\infty}^{\infty} \left[\exp \left\{ -\frac{(x-y+2na)^2}{4t} \right\} -\exp \left\{ -\frac{(x+y+2na)^2}{4t} \right\} \right].
\eeq
The conditional survival  probability is
\begin{align}
\Pr\{t_{1}>t\,|\,y \}=&\int\limits\limits_{0}^{a} p(x,t\,|\,y)\,dx\\
               =&\frac{1}{\sqrt{ 4 D \pi t}} \sum\limits_{n=-\infty}^{\infty}\int\limits\limits_{0}^{a}\left[\exp \left\{ -\frac{(x-y+2na)^2}{4t} \right\} -\exp \left\{ -\frac{(x+y+2na)^2}{4t} \right\} \right]\,dx\nonumber\\
               =& \int\limits\limits_{0}^{a}\frac{1}{\sqrt{ 4 D \pi t}}
               \left[ \exp \left\{ -\frac{(x-y)^2}{4t} \right\} -\exp \left\{ -\frac{(x+y)^2}{4t} \right\}\right]\,dx +S_1(y,t)-S_2(y,t),\nonumber
\end{align}
where 
\beqq
S_1&=&\frac{1}{\sqrt{ 4 D \pi t}} \sum\limits_{n=1}^{\infty}\int\limits\limits_{0}^{a}\left[\exp
\left\{ -\frac{(x+y+2na)^2}{4t} \right\} -\exp \left\{ -\frac{(x-y+2na)^2}{4t} \right\}
\right]\,dx\\
S_2&=&\frac{1}{\sqrt{ 4 D \pi t}} \sum\limits_{n=1}^{\infty}\int\limits\limits_{0}^{a}\left[\exp
\left\{ -\frac{(x+y-2na)^2}{4t} \right\} -\exp \left\{ -\frac{(x-y-2na)^2}{4t} \right\}
\right]\,dx.
\eeqq
Note that the integrand in the third line of (29), denoted $p_1(x,t\,|\,y)$, satisfies the
initial condition $p_1(x,0\,|\,y)=\delta(x-y)$ and the boundary condition
$p_1(0,t\,|\,y)=p_1(x,t\,|\,0)=0$, but $p_1(a,t\,|\,y)\neq0$ and $p_1(x,t\,|\,a)\neq0$.
However, with the first correction,
\begin{align}\label{2lines}
p_2(x,t\,|\,y)=&\,\frac{1}{\sqrt{ 4 D \pi t}}\left[ \exp \left\{ -\frac{(x-y)^2}{4t} \right\} -\exp \left\{ -\frac{(x+y)^2}{4t} \right\}\right.\\
+&\,\left.\exp \left\{ -\frac{(x-y-2a)^2}{4t} \right\} -\exp \left\{ -\frac{(x+y-2a)^2}{4t}
\right\}\right.\nonumber\\
+&\,\left.\exp \left\{ -\frac{(x-y+2a)^2}{4t} \right\} -\exp \left\{ -\frac{(x+y+2a)^2}{4t}
\right\}\right]\nonumber
\end{align}
{it} satisfies the same initial condition for $x$ and $y$ in the interval, and the boundary
conditions
\begin{align}
p_2(x,t\,|\,a)=0,\quad p_2(x,t\,|\,0)=\frac{1}{\sqrt{ 4 D \pi t}}\left[\exp\left\{-\frac{(x+2a)^2}{4t}\right\}-\exp\left\{-\frac{(x-2a)^2}{4t}\right\}\right].
\end{align}
Higher-order approximations correct the one boundary condition and corrupt the other, though the error decreases at higher exponential rates.
\begin{figure}[http!]
\centering
\includegraphics[width=.99\textwidth]{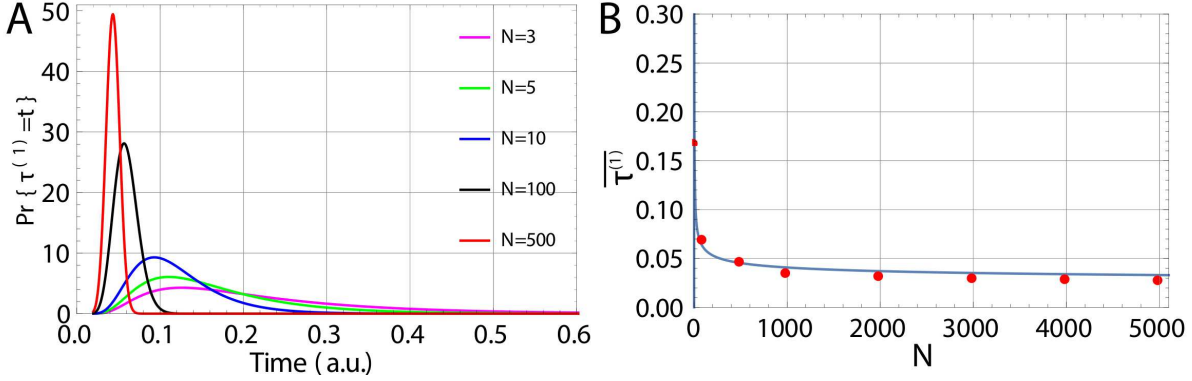}
\caption[caption]{\small{\textbf{A.}  Plot of $\Pr \{\tau^{(1)}=t\}$ (escape from an interval) for $N=3, 5, 10, 100$, and $500$ with $n_0= 100$ terms in the series of \eqref{eq:approx}. \textbf{B.} Decay of the expected arrival time of the fastest particle vs $N$ (red points). The plot of the asymptotic formula \eqref{taubar2N} is (blue) with parameter $\frac{0.282}{\log{N}}$}.} \label{fig:S1}
\end{figure}

The first line of \eqref{2lines} gives the approximation
\beq
&\ds\int\limits\limits_{0}^{a}\frac{1}{\sqrt{ 4 \pi t}} \left[\exp\left\{ -\frac{(x+y)^2}{4t} \right\} -\exp \left\{ -\frac{(x-y)^2}{4t} \right\}\right]\,dx
=\frac{1}{\sqrt{\pi}} \int\limits\limits_{(y-a)/2\sqrt{t}}^{y/2\sqrt{t}}e^{-u^2}\,du\nonumber\\
&\ds\sim1-\max\frac{2\sqrt{t}}{\sqrt{\pi}}\left[\frac{e^{-y^2/4t}}{y},\frac{e^{-(a-y)^2/4t}}{a-y}\right]
\quad\mbox{as}\ t\to0,
\eeq
where the maximum occurs at $\min[y,a-y]$ for $0<y<a$ (the shortest ray from $y$ to the boundary). Starting at $x=a/2$, this gives
\begin{align*}
 \Pr\{t_{1}>t\}= &\ds\int\limits\limits_{0}^{a}\frac{1}{\sqrt{ 4 \pi t}} \left[\exp \left\{ -\frac{(x-a/2)^2}{4t} \right\}-\exp\left\{ -\frac{(x+a/2)^2}{4t} \right\}\right]\,dx
\quad\mbox{as}\ t\to0,
\end{align*}
so changing  $x+a/2=z\sqrt{4t}$ in the first integral and $x-a/2=z\sqrt{4t}$ in the second, we get
\begin{align}\label{intgls}
\frac{1}{\sqrt{\pi}}\int\limits\limits_{-a/4\sqrt{t}}^{a/4\sqrt{t}}e^{-z^2}\,dz-
\frac{1}{\sqrt{\pi}}\int\limits\limits_{a/4\sqrt{t}}^{3a/4\sqrt{t}}e^{-z^2}\,dz
\approx&\,1-\frac{4\sqrt{t}e^{-a^2/16t}}{a\sqrt{\pi }}-\frac{2\sqrt{t}e^{-a^2/16t}}{a\sqrt{\pi }}-\frac{4\sqrt{t}e^{-9a^2/16t}}{6a\sqrt{\pi }}\nonumber\\
=&\,1-\frac{6\sqrt{t}e^{-a^2/16t}}{a\sqrt{\pi }}-\frac{2\sqrt{t}e^{-9a^2/16t}}{3a\sqrt{\pi }}.
\end{align}
{The second integral in the second line of \eqref{2lines}  is
\begin{align}\label{xa21}
I_{3/2}=-\ds\int\limits\limits_{0}^{a}\frac{1}{\sqrt{ 4 \pi t}} \left[\exp\left\{ -\frac{(x-3a/2)^2}{4t} \right\}\right]\,dx.
\end{align}
Set $x-3a/2=-z\sqrt{4t}$, then \eqref{xa21} becomes
\begin{align*}
I_{3/2}=-\ds\frac{1}{\sqrt{\pi}
}\int\limits\limits_{a/4\sqrt{t}}^{3a/4\sqrt{t}}\exp\left\{-z^2 \right\}\,dz=-\ds\frac{1}{\sqrt{\pi}
}\int\limits\limits_{a/4\sqrt{t}}^{\infty}\exp\left\{-z^2 \right\}\,dz+\ds\frac{1}{\sqrt{\pi}
}\int\limits\limits_{3a/4\sqrt{t}}^{\infty}\exp\left\{-z^2 \right\}\,dz.
\end{align*}
Thus the second line of \eqref{2lines} is
\begin{align}\label{xa2}
-&\ds\int\limits\limits_{0}^{a}\frac{1}{\sqrt{ 4 \pi t}} \left[\exp\left\{ -\frac{(x-3a/2)^2}{4t} \right\} -\exp \left\{ -\frac{(x-5a/2)^2}{4t} \right\}\right]\,dx\nonumber\\
\approx &-\frac{2\sqrt{t}}{a\sqrt{\pi}}e^{-a^2/16t}+\frac{2\sqrt{t}}{5a\sqrt{\pi}}e^{-25a^2/16t},
\end{align}
and in the third line of \eqref{2lines}, we get
\begin{align}\label{xa3}
&\ds\int\limits\limits_{0}^{a}\frac{1}{\sqrt{ 4 \pi t}} \left[\exp\left\{ -\frac{(x+3a/2)^2}{4t} \right\} -\exp \left\{ -\frac{(x+5a/2)^2}{4t} \right\}\right]\,dx\nonumber\\
\approx &\frac{2\sqrt{t}}{3a\sqrt{\pi}}e^{-9a^2/16t}-\frac{2\sqrt{t}}{5a\sqrt{\pi}}e^{-49a^2/16t},
\end{align}
hence
\beq
 \int\limits\limits_0 ^{\infty} \left[ \Pr\{t_{1}>t\} \right]^N dt \approx \int\limits\limits_0 ^{\infty} \exp\left\{ N\ln\left(1-\frac{8\sqrt{t}}{a\sqrt{\pi}}e^{-a^2/16t}\right) \right\}dt
\eeq
and the expected time of the fastest particle that starts at the center of the interval is \eqref{taubarN} with $a$ replaced by $a/2$ and $N$ replaced by $2N$. That is,
\beq
\bar{\tau}^{1} \approx \frac{a^2}{16D\ln\frac{2N}{\sqrt{\pi}}}\quad\mbox{for}\ N\gg1.\label{taubar2N}
\eeq}
Figure \ref{fig:S1}A shows plot of the pdf analytical approximation of shortest arrival time \eqref{eq:approx} with $n_0=100$ terms, $D=1$ and $L=1$ for  $N =4,6$, and $10$. As the number of particles increases, the mean first arrival time decreases (Fig.\ref{fig:S1}B) and according to equation \eqref{taubar2N}, the asymptotic behavior is given by $C/\log N,$ where $C$ is a constant. We show below the pdf of the fastest Brownian particle.
\section{The shortest NEP from a bounded domain in \font\bb=msbm10 at 18pt$\rR^{2,3}$}
To generalize the previous result to the case of $N$ i.i.d. Brownian particles in a bounded domain $\Omega\subset\rR^{2,3}$, we assume that the particles are
initially injected at a point $\y\in\Omega$ and they can escape through a single small absorbing window $\p\Omega_a$ in the boundary $\p\Omega$ of the domain. The pdf of the fist passage time to $\p\Omega_a$ is given by \eqref{arrv1}.
\subsection{Asymptotics in dimension 3}\label{sdim3}
To determine the short-time asymptotics of the pdf, we use the Laplace transform of the IBVP \eqref{IBVP} and solve the resulting elliptic mixed Neumann-Dirichlet BVP \cite{Narrow1}. The Dirichlet part of the boundary consists of $N$ well-separated small absorbing windows, $\p\Omega_a=\bigcup_{j=1}^N\p\Omega_{j}$ and the reflecting (Neumann) part is $\p\Omega_r=\p\Omega-\p\Omega_a$,  so that the IBVP \eqref{IBVP} has the form
\begin{align}\label{eqDP2}
\frac{\p p(\x,t\,|\,\y)}{\p t}=&D\Delta p(\x,t\,|\,\y)\\
p(\x,0\,|\,\y)= & \delta(\x-\y) \;\hbox{ for }\; \x,\y \in \Omega\nonumber \\
\ds \frac{\p p(\x,t\,|\,\y)}{\p \n} = & 0 \;\;\hbox{for}\;\; \x\, \in\, \p\Omega_r\nonumber\\
 p(\x,t\,|\,\y) = & 0 \;\;\hbox{ for }\;\;t>0,\ \x\, \in\, \p\Omega_a.\nonumber
 \end{align}
We consider the case $N=1$. The Laplace transform of \eqref{IBVP},
\beq
\hat p(\x,q\,|\,\y ) =\int\limits_0^{\infty} p(\x,t\,|\,\y)e^{-pt}\,dt,
\eeq
 gives the BVP
\begin{align*}
-\delta(\x-\y)+q\hat p(\x,q\,|\,\y )=&D\Delta \hat p(\x,q\,|\,\y )\hspace{0.5em}\mbox{for}\ \x,\y\in\Omega \\
\ds \frac{\p \hat p(\x,q\,|\,\y )}{\p \n} = & 0 \;\;\hbox{for}\;\; \x\, \in\, \p\Omega_r\\
\hat p(\x,q\,|\,\y )= & 0 \;\;\hbox{ for }\;\; \x\, \in\, \p\Omega_{a}.
 \end{align*}
Green's function for the Neumann problem in $\Omega$ is the solution of
\begin{align}
-\Delta_{\x} \hat G(\x,q\,|\,\y)+q\hat G(\x,q\,|\,\y) = & \delta(\x-\y)\hspace{0.5em} \hbox{ for } \x,\y\in \Omega, \label{eqDP2b}\\
\frac{\p \hat G_q(\x,q\,|\,\y)}{\p n_{\x}} = & 0\hspace{0.5em}\hbox{ for } \x,\y\in\p\Omega.\nonumber
\end{align}
The asymptotic solution of \eqref{eqDP2b} in $\rR^3$ is given by
\beq
\hat G(\x,q\,|\,\y)= \frac{e^{-\sqrt{q}|\x-\y|}}{4\pi||\x-\y||} +R_q(\x,\y),\label{PopoG}
\eeq
where $R_q(\x,\y)$ is more regular than the first term. When $\x$ (or $\y$) is on the boundary,
\beq
\hat G(\x,q\,|\,\y)= e^{-\sqrt{q}|\x-\y|}\left( \frac{1}{2\pi||\x-\y||}+ \frac{H(\x)}{2\pi}\log|\x-\y| +R(\x,\y)\right),
\eeq
where $R(\x,\y)$ is more regular than the logarithmic term and $H(\x)$ is a geometric factor \cite{Narrow1}. Using Green' identity, we obtain that
\begin{align*}
&\int\limits\limits_{\Omega} \left[\hat p(\x,q\,|\,\y ) \Delta_{\x} \hat G(\x,q\,|\,\y) - \Delta_{\x}\hat p(\x,q\,|\,\y ) \hat G(\x,q\,|\,\y) \right]\,d\y\\
=& \int\limits\limits_{\p \Omega} \left[ \hat p(\x,q\,|\,\y ) \frac{\p\hat G(\x,q\,|\,\y)}{\p n_{\x}} - \frac{\p \hat p(\x,q\,|\,\y )}{\p n_{\x}}\hat G(\x,q\,|\,\y)\right]\, dS_{\y},
\end{align*}
hence
\beq \label{represent}
\hat p(\x,q\,|\,\y )=  \hat G(\x,q\,|\,\y) -\int\limits\limits_{\p \Omega_a}  \frac{\p \hat p(\x,q\,|\,\y' )}{\p n_{\x}}\hat G(\x,q\,|\,\y')\,dS_{\y'}.
\eeq
If the absorbing window $\p \Omega_a $ is centered at  $\x =\A$, then, for $\x \in \p \Omega_a $,
\beq
0= \hat G(\A,q\,|\,\y) -\int\limits\limits_{\p \Omega_a} \frac{\p \hat p(\A,q\,|\,\y' )}{\p n_{\x}}\hat G(\A,q\,|\,\y')\, dS_{\y'}.
\eeq
This is a Helmoltz equation and the solution is given by {\cite{Narrow1}}
\beq
\hat p(\A,q\,|\,\y )=\frac{C}{\sqrt{a^2-r^2}},
\eeq
 where $r=|\A-\y|$. Thus, $C$ is computed from
\beq
0=  G_q(\A,\y) -\int\limits_{\p \Omega_a}  \frac{\p \hat p(\A,q,\y\,|\,\y )}{\p n_{\x}}G_q(\A,\y) dS_{\y}
\eeq
and  to leading order,
\beq
G_q(\A,\y)=\int\limits_{\p \Omega_a}  \frac{C e^{-\sqrt{q}|\y-\A|}}{\sqrt{a^2-r^2}}\left( \frac{1}{2\pi||\y-\A||}+ \frac{H(\x)}{2\pi}\log|\y-\A| +R(\y,\A) \right) dS_{\y}.
\eeq
 If $\p \Omega_a$ is a disk of radius $a$, then
\beq
G_q(\A,\y) \approx C \int\limits_{\p \Omega_a}  \frac{e^{-\sqrt{q}r}}{\sqrt{a^2-r^2}} \frac{1}{2\pi r} 2\pi r dr= \frac{\pi}{2} \left(I_0(\sqrt{q}a)-L_0(\sqrt{q}a)\right)C,
\eeq
where $I_0$ is the modified Bessel function of the first kind and $L_0$ the Struve function.
Thus,
\beq
\hat p(\x,q\,|\,\y )=  G_q(\x,\y) - G_q(\A,\y)\frac{2}{\pi \left(I_0(\sqrt{q}a)-L_0(\sqrt{q}a) \right)} \int\limits_{\p \Omega_a}  \frac{G_q(\x,\y) dS_{\y}}{\sqrt{a^2-r^2}} .
\eeq
For $|\A-\x|\gg a$ and q large, $I_0(\sqrt{q}a)-L_0(\sqrt{q}a)\approx \frac{2}{\pi \sqrt{q}a}$
\beqq
\hat p(\x,q\,|\,\y )\approx G_q(\x,\y) - G_q(\A,\y) G_q(\A,\x)\frac{2}{\pi (I_0(\sqrt{q}a)-L_0(\sqrt{q}a))} \int\limits_{\p \Omega_a}  \frac{dS_{\y}}{\sqrt{a^2-r^2}} ,
\eeqq
hence, for a small circular window of  radius $a$
\beq
\hat p(\x,q\,|\,\y )\approx   G_q(\x,\y) - 2\pi  \sqrt{q}a^2 G_q(\A,\y)G_q(\A,\x) +o(a^2)\quad\mbox{for}\ a\ll1.
\eeq
To leading order in small $t$ and $\x,\y \in \Omega$, we obtain for the first term using the leading order term in expression \ref{PopoG},
\beq
{\mathcal L}^{-1}(G_q(\x,\y))\approx\ds\frac{1}{{(4\pi t)^{3/2}}} e^{\ds -\frac{|\x-\y|^2}{4t}}.
\eeq
We will now use the inverse Laplace transform \cite{Abramowitz}[p.1026; 29.3.87]
\beq
{\mathcal L}^{-1}(\sqrt{q}\frac{e^{-\sqrt{q}|\x-\y|}}{|\x-\y|}) = \ds \frac{1}{4\sqrt{\pi t^3}} e^{\ds -\frac{|\x-\y|^2}{4t}}H_2(\frac{|\x-\y|}{2\sqrt{t}}),
\eeq
where $H_2(x)=4x^2-2$ is the Hermite polynomial of degree 2.  With $\A \in \p \Omega$, the image charge (for the Dirichlet boundary) leads to a factor $1/2$ and we write
\beq
G_q(\A,\y)G_q(\A,\x)2\pi  \sqrt{q}a^2 = \sqrt{q}a^2 \frac{e^{\ds-\sqrt{q}(|\A-\y|+|\A-\x|)}}{ 2 \pi|\A-\y||\A-\x|}
\eeq
and
\beq
{\mathcal L}^{-1}(G_q(\A,\y)G_q(\A,\x)\sqrt{q}a^2)= \frac{a^2}{(4\pi t)^3} \frac{e^{-\ds\frac{(|\A-\y|+|\A-\x|)^2}{4t}}}{{|\A-\y||\A-\x|}}H_2(\frac{|\A-\y|+|\A-\x|}{2\sqrt{t}}).
\eeq
Finally,
{\small
\beq
\ds {\mathcal L}^{-1}(\hat p(\x,q\,|\,\y ))=\frac{1}{\sqrt{(4\pi t)^3}} \left(
e^{\ds -\frac{|\x-\y|^2}{4t}}- \frac{a^2}{{|\A-\y||\A-\x|}} e^{-\ds\frac{(|\A-\y|+|\A-\x|)^2}{4t}}H_2(\frac{|\A-\y|+|\A-\x|}{2\sqrt{t}})\right).
\eeq
}
The short-time asymptotics of the survival probability with $\delta=|\A-\y|$,
{\small
\beq
S(t)&\approx&\ds\int\limits_{\Omega} p_t(\x,\y ) d\x \\
&=& \frac{1}{\sqrt{(4\pi t)^3}}   \int\limits_{\Omega}\left(
e^{-\ds\frac{|\x-\y|^2}{4t}}- \frac{a^2}{{|\A-\y||\A-\x|}} e^{-\ds\frac{(|\A-\y|+|\A-\x|)^2}{4t}}H_2(\frac{|\A-\y|+|\A-\x|}{2\sqrt{t}}) \right)d\x \nonumber\\
 &=&\ds I_1(t)-I_2(t)-I_3(t)-I_4(t),
\eeq
}
where for small t,
\beq
H_2(\frac{|\A-\y|+|\A-\x|}{2\sqrt{t}}) \approx \frac{(|\A-\y|+|\A-\x|)^2}{t},
\eeq
and
\beq
I_1(t)&=& \ds \frac{1}{\sqrt{(4\pi t)^3}}  \int\limits_{\Omega}
e^{-\ds\frac{|\x-\y|^2}{4t}}d\x   \\
I_2(t)&=& \frac{1}{\sqrt{(4\pi t)^3}} \frac{a^2\delta}{t} \int\limits_{\Omega}  \frac{1}{{|\A-\x|}}
e^{-\ds\frac{(\delta+|\A-\x|)^2}{4t}}d\x        \\
I_3(t)&=&  \frac{1}{\sqrt{(4\pi t)^3}}\frac{2a^2}{t} \int\limits_{\Omega}
e^{-\ds\frac{(\delta+|\A-\x|)^2}{4t}}d\x. \\
I_4(t)&=&\frac{1}{\sqrt{(4\pi t)^3}}\frac{a^2}{\delta t} \int\limits_{\Omega} |\A-\x|
e^{-\ds\frac{(\delta+|\A-\x|)^2}{4t}}d\x.
\eeq
 Each integral is evaluated in the short-time limit.
\beq\label{r1}
I_1(t)&=& \ds \frac{1}{\sqrt{(4\pi t)^3}}  \int\limits_{\Omega}
e^{-\ds\frac{|\x-\y|^2}{4t}}d\x \approx1-\frac{2}{\sqrt{\pi}} \int\limits\limits_{R_a/\sqrt{4t}}^{\infty}e^{-u^2}\,du\\
& \approx& 1-\sqrt{4t}\frac{e^{-\left(R_a/\sqrt{4t}\right)^2}}{R_a\sqrt{\pi}} \left(1+O\left(\left(R_a/\sqrt{4t}\right)^2\right)\right),
\eeq
where $R_a$ is the radius of the maximal ball  inscribed in $\Omega$. The integral $I_2$ is evaluated by the change of variables $\z=\x -\A$ and then $\eta =(\delta+r)/\sqrt{4t}$, where $r=|\z|$ {(recall that $\A$ is in $\Omega_a$)},
\beq
I_2(t)&=& \frac{1}{\sqrt{(4\pi t)^3}}  \frac{a^2\delta}{t}\int\limits_{\Omega}  \frac{1}{{|\A-\x|}}
e^{-\ds\frac{(\delta+|\A-\x|)^2}{4t}}d\x   \\
      &=& \frac{1}{\sqrt{(4\pi t)^3}} \frac{a^2\delta}{t}\int\limits_{\Omega+\A} \frac{1}{{|\z|}}
e^{-\ds\frac{(\delta+|\z|)^2}{4t}}2\pi |\z|^2d|\z|  \\
 &\approx& \frac{2\pi }{\sqrt{(4\pi t)^3}} \frac{a^2\delta}{t} \int\limits_{\frac{\delta}{\sqrt{4t}}}^{\frac{\delta+R}{\sqrt{4t}}}
e^{-\ds \eta^2} (\sqrt{4t}\eta-\delta) \sqrt{4t}d\eta,
\eeq
where $R$ is the radius of the largest half-ball centered at $\A\in\Omega_a$ and inscribed in $\Omega$. For short time,
\beq \label{gaussianint}
\int\limits_{\frac{\delta}{\sqrt{4t}}}^{\frac{\delta+R}{\sqrt{4t}}} e^{-\ds \eta^2}d\eta&\approx&\frac{1}{2}
\left\{ \frac{\sqrt{4t}}{\delta}e^{-\ds \left(\frac{\delta}{\sqrt{4t}}\right)^2} \right\} \left(1-\frac{4t}{2\delta^2}+12\frac{t^2}{\delta^4} \right)
\\
\int\limits_{\frac{\delta}{\sqrt{4t}}}^{\frac{\delta+R}{\sqrt{4t}}} \eta e^{-\ds \eta^2}d\eta&=& \frac12\left( e^{-\ds \left(\frac{\delta}{\sqrt{4t}}\right)^2}-e^{-\ds \left(\frac{\delta+R}{\sqrt{4t}}\right)^2}\right)\approx \frac{1}{2}e^{-\ds \left(\frac{\delta}{\sqrt{4t}}\right)^2}.
\eeq
Therefore,
{\small
\beq \label{r2}
I_2(t)&\approx& \frac{1}{\sqrt{(4\pi t)^3}}\frac{2\pi a^2\delta}{t} \left( 2t e^{-\ds \left(\frac{\delta}{\sqrt{4t}}\right)^2}-\delta \sqrt{4 t} \frac{1}{2}
\left( \frac{\sqrt{4t}}{\delta}e^{-\ds (\frac{\delta}{\sqrt{4t}})^2} \right)(1-\frac{4t}{2\delta^2}+12\frac{t^2}{\delta^4})  \right)\\
&\approx& \frac{4a^2}{\delta \sqrt{\pi}}\frac{1}{\sqrt{t}}(1-\frac{6t}{\delta^2})e^{-\ds \left(\frac{\delta}{\sqrt{4t}}\right)^2}.
\eeq
}
Now,
\beq
I_3(t)&=& \frac{1}{\sqrt{(4\pi t)^3}} \frac{2a^2}{t} \int\limits_{\Omega}
e^{-\ds\frac{(\delta+|\A-\x|)^2}{4t}}d\x \\
&=& \frac{2\pi}{\sqrt{(4\pi t)^3}}\frac{2a^2}{t} \int\limits_{\frac{\delta}{\sqrt{4t}}}^{\frac{\delta+R}{\sqrt{4t}}}
e^{-\ds \eta^2} (\sqrt{4t}\eta-\delta)^2 \sqrt{4t}d\eta \\
&=& \frac{2\pi}{\sqrt{(4\pi t)^3}} \frac{2a^2}{t} \int\limits_{\frac{\delta}{\sqrt{4t}}}^{\frac{\delta+R}{\sqrt{4t}}}
e^{-\ds \eta^2} (4t\eta^2-2\sqrt{4t}\eta \delta +\delta^2)\sqrt{4t}d\eta,
\eeq
that we write as $I_3(t)=I^{(1)}_3(t)+I^{(2)}_3(t)+I^{(3)}_3(t)$.  The approximation
\beq
\int\limits_{\frac{\delta}{\sqrt{4t}}}^{\frac{\delta+R}{\sqrt{4t}}}
e^{-\ds \eta^2} \eta^2 d\eta \approx \frac{\delta}{2\sqrt{4t}}\, e^{-\ds \left(\frac{\delta}{\sqrt{4t}}\right)^2} + \frac{1}{4}
\left( \frac{\sqrt{4t}}{\delta}e^{-\ds \left(\frac{\delta}{\sqrt{4t}}\right)^2} \right) \left(1-\frac{4t}{2\delta^2}+ o(t)\right),
\eeq
gives
\begin{align*}
I^{(1)}_3(t)=&\frac{2 \pi \delta}{\pi^{3/2}} \frac{a^2}{{t \sqrt{4t}}}e^{-\ds \{\frac{\delta}{\sqrt{4t}}\}^2}+
\frac{2a^2}{2t\pi^{3/2}} \frac{1}{{\delta^2}}
 \sqrt{4t}e^{-\ds (\frac{\delta}{\sqrt{4t}})^2}  \left(1-\frac{4t}{2\delta^2}\right)\\
I^{(2)}_3(t)=&-\frac{ 4\pi a^2\delta}{t\sqrt{4t} \pi^{3/2}}e^{-\ds (\frac{\delta}{\sqrt{4t}})^2}\\
I^{(3)}_3(t)=&\frac{ 2\pi a^2\delta}{t\sqrt{4t} \pi^{3/2}}
 e^{-\ds \left(\frac{\delta}{\sqrt{4t}}\right)^2} \left(1-\frac{4t}{2\delta^2}+o(t)\right).
\end{align*}
Summing the three contributions, the leading order terms cancel and
\beq \label{r3}
I_3(t)=\frac{4a^2\sqrt{t}}{\pi^{1/2}\delta^3} e^{-\ds \left(\frac{\delta}{\sqrt{4t}}\right)^2}.
\eeq
To compute $I_4$, we decompose it into 4 pieces:
\beq \label{r4}
I_4(t)&=&\frac{1}{\sqrt{(4\pi t)^3}}\frac{a^2}{\delta t} \int\limits_{\Omega} |\A-\x|
e^{-\ds\frac{(\delta+|\A-\x|)^2}{4t}}d\x.\\
      &=& \frac{2\pi}{\sqrt{(4\pi t)^3}}\frac{a^2}{\delta t} \int\limits_{\frac{\delta}{\sqrt{4t}}}^{\frac{\delta+R}{\sqrt{4t}}}
e^{-\ds \eta^2} (\sqrt{4t}\eta-\delta)^3 \sqrt{4t}d\eta \\
&=& J_1(t)+J_2(t)+J_3(t)+J_4(t).
\eeq
Direct computations give:
\beq
J_1(t)=\frac{2\pi(4t)^2}{\sqrt{(4\pi t)^3}}\frac{a^2}{\delta t} \int\limits_{\frac{\delta}{\sqrt{4t}}}^{\frac{\delta+R}{\sqrt{4t}}}
e^{-\ds \eta^2}\eta^3 d\eta=   \frac{4a^2 \delta}{\sqrt{\pi} (4t)^{3/2}} (1+\frac{4t}{\delta^2}) e^{-\ds \left(\frac{\delta}{\sqrt{4t}}\right)^2},
\eeq
where we used
\beq
\int\limits_{\frac{\delta}{\sqrt{4t}}}^{\frac{\delta+R}{\sqrt{4t}}} \eta^3 e^{-\ds \eta^2}d\eta&\approx& \left( 1+\frac{\delta^2}{4t}\right) e^{-\ds \left(\frac{\delta}{\sqrt{4t}}\right)^2}.
\eeq
Next,
{\small
\beq
J_2(t)=-\frac{2\pi}{\sqrt{(\pi)^3}}\frac{a^2}{\delta t} \int\limits_{\frac{\delta}{\sqrt{4t}}}^{\frac{\delta+R}{\sqrt{4t}}}
e^{-\ds \eta^2} 3\eta^2 \delta d\eta=-\frac{12a^2\delta}{ \sqrt{\pi}(4t)^{3/2}}\left(1+\frac{2t}{\delta^2}(1-\frac{2t}{\delta^2}+\frac{12t^2}{\delta^4}+o(t^2)) \right)e^{-\ds \left(\frac{\delta}{\sqrt{4t}}\right)^2},
\eeq
}
where we have used
\beq
\int\limits_{\frac{\delta}{\sqrt{4t}}}^{\frac{\delta+R}{\sqrt{4t}}} \eta^2 e^{-\ds \eta^2}d\eta&\approx& \left( \frac{\delta}{2\sqrt{4t}}
+\frac{\sqrt{4t}}{4\delta}\left(1-\frac{4t}{2\delta^2}+12\frac{t^2}{\delta^4} +o(t^2) \right)\right)e^{-\ds \left(\frac{\delta}{\sqrt{4t}}\right)^2}
\eeq
Using relation \ref{gaussianint},
\beq
J_3(t)=\frac{2\pi}{\sqrt{(4t\pi^3)}}\frac{a^2}{\delta t} \int\limits_{\frac{\delta}{\sqrt{4t}}}^{\frac{\delta+R}{\sqrt{4t}}}
e^{-\ds \eta^2} 3\eta \delta^2 d\eta= \frac{12a^2\delta}{\sqrt{\pi}(4t)^{3/2}}e^{\ds -\frac{\delta^2}{4t}}.
\eeq
Finally,
\beq
J_4(t)=-\frac{2\pi}{4t\sqrt{(\pi)^3}}\frac{a^2}{\delta t} \int\limits_{\frac{\delta}{\sqrt{4t}}}^{\frac{\delta+R}{\sqrt{4t}}}
e^{-\ds \eta^2} \delta^3 d\eta= \ds -\frac{4 a^2 \delta}{(4t)^{3/2}\sqrt{\pi}}e^{\ds -\frac{\delta^2}{4t}}\left( 1-\frac{2t}{\delta^2} +\frac{12t^2}{\delta^4}+o(t^2)\right).
\eeq
Direct computations show that the terms in $t^{-3/2}$ and $t^{-1/2}$ cancels out in the computation of $I_4$ from the four terms $J_1,..J_4$ and it remains only the term in $t^{1/2}$
\beq \label{r4f}
I_4(t)=-\frac{9a^2}{\sqrt{\pi}\delta^3} t^{1/2} e^{\{-\delta^2/4t\}}.
\eeq
Summing \eqref{r1}-\eqref{r3}-\eqref{r4}, we get
\beqq
S(t)&=&\ds\int\limits_{\Omega} p_t(\x,\y ) d\x \\
&=&\ds 1-\sqrt{4t}\,\frac{e^{-\left(R_a/\sqrt{4t}\right)^2}}{R_a\sqrt{\pi}}
 -\frac{a^2}{\delta \pi^{1/2}\sqrt{t}}\,e^{-\delta^2/4t} +o\left(t^{1/2}e^{-\ds \left(\frac{\delta}{\sqrt{4t}}\right)^2}\right)\\
 &\approx & \ds 1-\frac{a^2}{\delta \pi^{1/2}\sqrt{t}}\,e^{-\delta^2/4t}. \nonumber
\eeqq
It follows that  in three dimensions, the expected shortest arrival time to a small circular window of radius $a$, the expected shortest time $\bar{\tau}^{3}$ is given by
\beqq
\bar{\tau}^{3}=  \int\limits\limits_0 ^{\infty} \left[ \Pr\{t_{1}>t\} \right]^N dt &\approx&
 \int\limits\limits_0 ^{\infty} \exp N\log\left(1 -\frac{a^2}{\delta \pi^{1/2}\sqrt{t}}\,e^{-\delta^2/4t}  \right)dt \\
  &\approx& \int\limits\limits_0 ^{\infty} \exp \left( -N\frac{4(a/\delta) }{\delta \pi^{3/2}  \sqrt{t}}e^{-\delta^2/4t} \right) dt\\
   &\approx& \delta^2 \int\limits\limits_0 ^{\infty} \exp \left( -N' \frac{1}{\sqrt{u}}e^{-1/4u} \right) du,
\eeqq
where $N'=N\frac{4a^2}{\pi^{1/2}\delta^2}$
Using the method develop in section \ref{rayescape} with the change of variable,
\beq
w=w(t)=\frac{1}{\sqrt{t}}e^{-1/t}, \quad
w'(t)=\frac{1}{\sqrt{t}}e^{-\frac{1}{t}}\left(-\frac{1}{2t} +\frac{1}{4t^{3/2}}\right).
\eeq
We have with $w'=4w(\log(w))^{3/2}$
\beqq
\bar{\tau}^{3} &\approx& \delta^2 \int\limits\limits_0 ^{\infty} \frac{\exp (-N'w)}{4w(\log(w))^{3/2}} du
\eeqq
When the diffusion coefficient is $D$, the formula changes to
\beq \label{taudim3}
\bar{\tau}^{3}\approx \frac{\delta^2}{2D\sqrt{\log\left(\ds N\frac{4a^2}{\pi^{1/2}\delta^2}\right)}}.
\eeq
The next term in the expansion can be obtained by  accounting for the logarithmic singularity in the expansion of Green's function. When there are $p$ windows, whose distances from the initial position of the Brownian particle are $d_k=dist(P_0,P_k $), formula \eqref{taudim3} changes to {
\beq
\bar{\tau}^{3}\approx \frac{\delta^2_m}{2D\sqrt{\log\left(\ds N\frac{4a^2}{\pi^{1/2}\delta^2}\right)}},
\eeq
where $\delta^2_m=\min(d_1^2,..d_p^2)$. The asymptotic formula \eqref{taudim3} is compared with results of Brownian simulations and shows very good agreement (Fig. \ref{fig:5}). When absorbing windows are ellipses,  the Green's function approach, based on Narrow Escape methodology, can be applied as well \cite{holcmanschuss2015}.
\begin{figure}[http!]
\centering
\includegraphics[width=.99\textwidth]{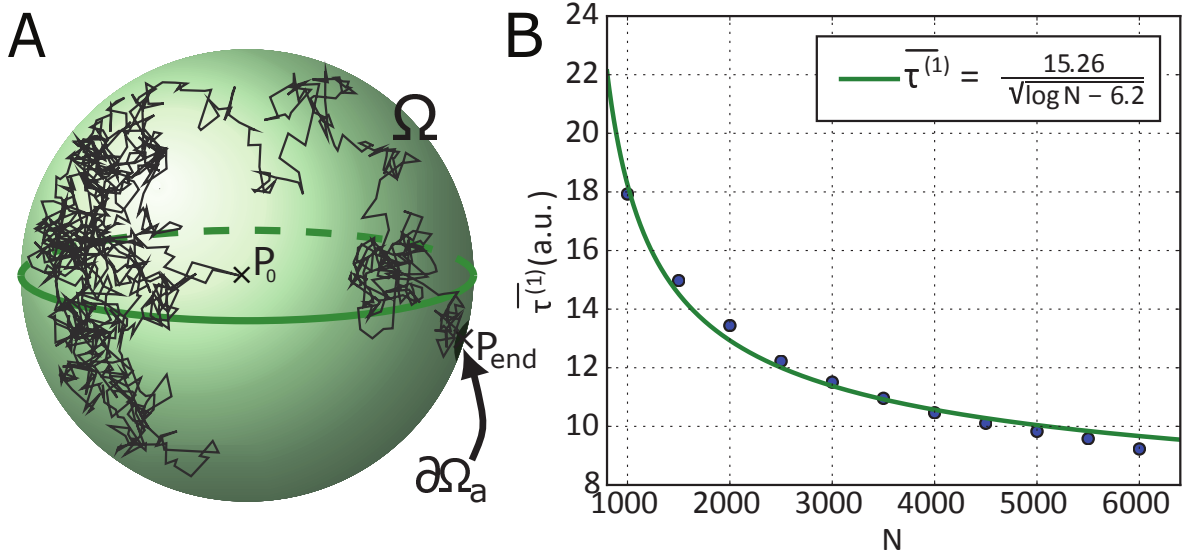}
\caption[caption]{\small Extreme statistics of the narrow escape time through a small window in three dimensions. \textbf{A.} The geometry  of the NEP for the fastest particle. In the simulation, the sphere has a radius $5\, \mu m$, the absorbing window  $\partial {\mathcal{S}}_a $, has a radius $\varepsilon = 0.1\mu m$ and the diffusion coefficient is $D = 0.2\mu m^2 s^{-1}$. The trajectory starts at point $P_0$ (cross), and ends at point $P_{end}$ . \textbf{B.} Plot of the MFPT of the fastest particle versus the number of particle $N$. We simulated 2000 runs. The asymptotic solution (red curve) is $A/\log( N+B)$.}\label{fig:5}
\end{figure}

\subsection{Asymptotics in dimension 2}
We consider the diffusion of $N$ Brownian i.i.d. particles in a two-dimensional
domain $\Omega$  with a small absorbing arc $\p\Omega_a$ of length $2\eps$ on the otherwise
reflecting boundary $\p\Omega$. To compute the pdf of the shortest arrival time to the arc, we follow the
steps of the analysis in dimension 3, presented in the previous subsection  \ref{sdim3}.

The Neumann-Green function \eqref{eqDP2b} in two dimensions is the solution of the BVP
\begin{align}
-\Delta_{\x} \hat G(\x,q\,|\,\y)+q\hat G(\x,q\,|\,\y) = & \delta(\x-\y)\hspace{0.5em} \hbox{ for } \x,\y\in \Omega, \label{eqDP2b3}\\
\frac{\p \hat G_q(\x,q\,|\,\y)}{\p n_{\x}} = & 0\hspace{0.5em}\hbox{ for } \x,\y\in\p\Omega,
\end{align}
is given for $\x,\y\in\p\Omega$ by \cite[p.51]{Ward2011}
\begin{align} \label{Gexpression}
\hat G(\x,q\,|\,\y)=\frac{1}{\pi}K_0(\sqrt{q}|\x-\y|)+R(\x,\y),
\end{align}
where $R(\x,\y)$ is its regular part. For a disk, the analytical expression is given by the
series \begin{align}
R(\x,\y) =\frac{1}{\pi} \sum_{0}^{\infty} \sigma_n \cos(n(\psi-\psi_0))\frac{K'_n(\sqrt{q})}{I'_n(\sqrt{q})}I_n(r\sqrt{q})I_n(r_0\sqrt{q}),
\end{align}
where $\sigma_0=1, \sigma_n=2$ for $n\geq 2$ and $\x=re^{i\psi},\y=r_0e^{i\psi_0}$. The integral representation \eqref{represent} of the solution is
\beq \label{represent2}
\hat p(\x,q\,|\,\y )=  \hat G(\x,q\,|\,\y) -\int\limits\limits_{\p \Omega_a}  \frac{\p \hat p(\x,q\,|\,\y' )}{\p n_{\x}}\hat G(\x,q\,|\,\y')\,dS_{\y'},
\eeq{
so choosing $\x \in \p \Omega_a $,
\beq
0= \hat G(\x,q\,|\,\y) -\int\limits\limits_{\p \Omega_a} \frac{\p \hat p(\x,q\,|\,\y' )}{\p n_{\x}}\hat G(\x,q\,|\,\y')\, dS_{\y'}.
\eeq
This Helmholtz equation has the constant solution \cite{holcmanschuss2015}
\beq
\frac{\p \hat p(\x,q\,|\,\y' )}{\p n_{\x}}=C\hspace{0.5em}\mbox{for all}\ \x=\A\in\p\Omega_a.
\eeq
To leading order, we get
\beq
\hat G(\A,q\,|\,\y)=\frac{C}{\pi}\int\limits_{\p \Omega_a}K_0(\sqrt{q}|\A-\y|)\, ds_{\y},
\eeq
where $ds_{\y}$ is arclength element in $\p\Omega_a$.
When $|\x-\y|\leq 4 \eps$ and $\sqrt{q}\eps \ll 1$ in the large $q$ expansion of Green's function,
\beq
K_0(\sqrt{q}|\x-\y|) =-\log(\sqrt{q}|\x-\y|)+\log 2 -\gamma_0+o(1),
\eeq
we obtain
\beq
\hat G(\A,q\,|\,\y)=\frac{C}{\pi}\int\limits_{\p \Omega_a}   \left[ \ds -\log(\sqrt{q}|\A-\y|)+\log 2 -\gamma_0+o(1)\right]\, ds_{\y};
\eeq
that is,
\beq
\hat G(\A,q\,|\,\y)=\frac{2C}{\pi}\int\limits_{0}^{\eps}\left[ \ds -\log(\sqrt{q}r)+\log 2 -\gamma_0+o(1)\right]\, dr.
\eeq
Therefore the leading order approximation of $C$ is
\beq
C = \frac{\pi\hat G(\A,q\,|\,\y)}{2\eps \left[ \ds -\log(\sqrt{q}\eps)+O(\eps)\right]}.
\eeq
Finally, \eqref{represent2}  gives for $|\A-\x|\gg \eps$
\beq
\hat p(\x,q\,|\,\y )\approx \hat G(\x,q\,|\,\y) + \frac{\pi\hat G(\A,q\,|\,\y)\hat G(\A,q\,|\,\x) }{\ds \log(\sqrt{q}\eps)+O(\eps)}.
\eeq

The inversion formula  \cite[p.1028]{Abramowitz} for $k>0$,
\beq
{\mathcal L}^{-1}(K_0(k\sqrt{q}) = \ds \frac{1}{2t} e^{\ds -\frac{k^2}{4t}},
\eeq
gives
\beq
{\mathcal L}^{-1} \hat G(\x,q\,|\,\y)=\ds\frac{1}{4\pi t}
e^{-\ds\frac{|\x-\y|^2}{4t}}.
\eeq
For an initial point far from the boundary layer near the window, the expansion \cite[p.378]{Abramowitz}
\beq
K_0(z) = \ds \sqrt{\frac{\pi}{2z}} e^{\ds -z} \left(1+O\left(\frac1{z}\right)\right)\quad\mbox{for}\ z\gg1,
\eeq
gives in \eqref{Gexpression}
\beq
\hat G(\A,q\,|\,\y)\hat G(\A,q\,|\,\x) =\ds \frac1{2\pi}\sqrt{\frac{1}{q{s_1 s_2}}} e^{\ds -\sqrt{q}(s_1+s_2)} \left(1+O(q^{-1/2})\right),
\eeq
where $s_1=|\A-\y|$ and $s_2=|\A-\x|$,
\beqq
 \frac{\pi \hat G(\A,q\,|\,\y)\hat G(\A,q\,|\,\x)}{\ds -\log(\sqrt{q}\eps)+O(\eps)}&=&\ds \frac1{-2\log(\sqrt{q}\eps)}\sqrt{\frac{1}{q{s_1 s_2}}} e^{\ds -\sqrt{q}(s_1+s_2)} \left(1+O(q^{-1/2})\right),\\
&\approx & \ds \frac1{-2\log(\eps) +O(1)}\sqrt{\frac{1}{q{s_1 s_2}}} e^{\ds -\sqrt{q}(s_1+s_2)} \left(1+O(q^{-1/2})\right).
\eeqq
The inversion formula
\beq
{\mathcal L}^{-1}\left(\frac{1}{\sqrt{q}}e^{\ds -k\sqrt{q}}\right) = \ds \frac1{\sqrt{\pi t}} e^{\ds -\frac{k^2}{4t}}
\eeq
gives
\beq
 \frac{\pi {\mathcal L}^{-1}(\hat G(\A,q\,|\,\y)\hat G(\A,q\,|\,\x)}{ \ds -\log(\sqrt{q}\eps)+O(\eps)})=
\frac1{-2\log(\eps) +O(1)} \ds \frac1{\sqrt{\pi ts_1 s_2}} e^{\ds -\frac{(s_1+s_2)^2}{4t}}.
\eeq
Hence, we obtain the short-time asymptotics of the survival probability
\beq \label{survivdim2}
S(t)&\approx&\ds\int\limits_{\Omega} p_t(\x,\y )\, d\x \\
&=&\ds \frac{1}{4\pi t}   \int\limits_{\Omega}
e^{-\ds\frac{|\x-\y|^2}{4t}}\,d\x \\&&-\frac1{-2\log(\eps)\sqrt{s_2} +O(1)}\frac1{\sqrt{\pi t}}\int\limits_{\Omega}\sqrt{\frac{1}{{|\A-\x| }}}  \ds e^{\ds -\frac{(|\A-\x|+s_2)^2}{4t}}d\x\\
&=&R_1(t)+R_2(t)
\eeq
where
\beq\label{r1-dim2}
R_1(t)&=& \ds \frac{1}{4\pi t}  \int\limits_{\Omega}
e^{-\ds\frac{|\x-\y|^2}{4t}}d\x \approx 1-{e^{-\left(R_a/\sqrt{4t}\right)^2}},
\eeq
and $R_a$ is the radius of the maximal disk  inscribed in $\Omega$. The second term is
\beq\label{r2-dim2}
R_2(t)&=& -\frac1{-2\log(\eps)\sqrt{s_2} +O(1)}\frac1{\sqrt{\pi t}}\int\limits_{\Omega}\sqrt{\frac{1}{{|\A-\x| }}}  \ds\, e^{\ds -\frac{(|\A-\x|+s_2)^2}{4t}}d\x\nonumber \\
&\approx& -\frac1{-2\log(\eps)\sqrt{s_2} +O(1)}\sqrt{\frac{\pi}{ t}}\int\limits_{0}^{R_a} \ds e^{\ds -\frac{(r+s_2)^2}{4t}}\sqrt{r}\,dr.
\eeq
The small $t$ Laplace expansion and the two successive changes of variable $u=\frac{s_2}{2t}r$ and $v=u^{3/2}$ give
\beq
\int\limits_{0}^{R_a} \ds e^{\ds -\frac{(r+s_2)^2}{4t}}\sqrt{r}dr \approx  \frac{2}{3} \left(\frac{2t}{s_2}\right)^{3/2}e^{\ds -\frac{s_2^2}{4t}} \int\limits_0^{\infty} e^{-v^{3/2}}dv.
\eeq
Thus, with
\beq
I=\int\limits_0^{\infty} e^{-v^{2/3}}dv =\frac{3\sqrt{\pi}}{4},
\eeq
we obtain
\beq\label{r2b-dim2}
R_2(t)  -\frac1{-2\log(\eps) +O(1)}\frac{ \sqrt{2} \pi t }{s_2^2} \ds e^{\ds -\frac{s_2^2}{4t}}.
\eeq
We conclude therefore that the survival probability \eqref{survivdim2} is approximately
\beq
S(t)&\approx&1-\frac1{2\log(\frac{1}{\eps})}\frac{ \sqrt{2} \pi t }{s_2^2} \ds e^{\ds -\frac{s_2^2}{4t}},
\eeq
where the contribution of \eqref{r1-dim2} is negligible. Thus the MFPT of the fastest particle is
given by
\beqq
\bar{\tau}^{2}=  \int\limits\limits_0 ^{\infty} \left[ \Pr\{t_{1}>t\} \right]^N dt &\approx&
 \int\limits\limits_0 ^{\infty} \exp\left\{ N\log\left(1-\frac1{2\log(\frac{1}{\eps})}\frac{ \sqrt{2} \pi t }{s_2^2} \ds e^{\ds -\frac{s_2^2}{4t}}
 \right) \right\}dt \\
  &\approx& \int\limits\limits_0 ^{\infty} \exp \left\{ -N\frac1{2\log(\frac{1}{\eps})}\frac{ \sqrt{2} \pi t }{s_2^2} \ds e^{\ds -\frac{s_2^2}{4t}}\right\} dt.
\eeqq
The computation of the last integral follows the steps described in subsection \ref{s:rayescape}. The change of variable $w=te^{\ds -\frac{s_2^2}{4t}}$ leads to the asymptotic formula with diffusion coefficient $D$
\beqq
\bar{\tau}^{2}\approx \ds \frac{\ds s_2^2}{\ds 4 D \log\left(\frac{\pi \sqrt{2}N}{8\log\left(\frac{1}{\eps}\right)}\right)}.
\eeqq
where $s_2=|\x-\A|$ and $\x$ is the position of injection $\A$ the center of the absorbing window. This formula is compared with  Brownian simulations in Fig. \ref{fig:5b}. However, the dependence on the window size is $\log (\frac{1}{\eps})$.}
\begin{figure}[http!]
\centering
\includegraphics[width=.79\textwidth]{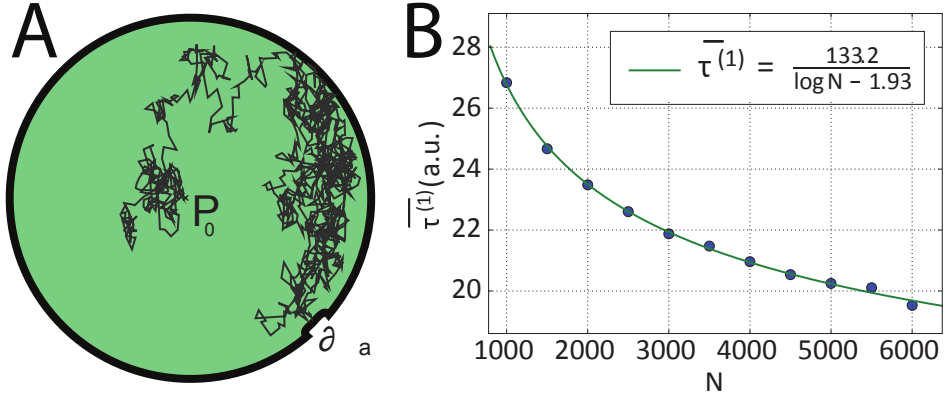}
\caption[caption]{\small Escape through a narrow opening in a planar disk. \textbf{A.}
The geometry of the NEP for the fastest particle. \textbf{B.} Plot of the MFPT of the fastest
particle versus the number of particles $N$. The asymptotic solution (red curve) is  of the form $\frac{\alpha}{\log (N)+\beta}$.}\label{fig:5b}
\end{figure}
\section{ Application of extreme statistics in cell biology}
\subsection{The first arrival time of ions in a dendritic-spine geometry}
The geometry of a dendritic spine (see Fig.\ref{fig:spinefig4}) is composed of a
head with a small hole opening, connected to a cylindrical neck. Initially, all Brownian
particles, which represent calcium ions that are uniformly distributed in the spine head at the
time of their release. This geometry  implies that the mean time $\tau$ to reach the base of the
neck is the mean time $\tau^{1}$ to reach the small window for the first time plus the  mean time $\tau^{2}$ spent in the spine neck, with no possible returns: we assume here that when a particle enters the neck cylinder, it cannot return to the head.

We compute now the distribution of the arrival time for the fastest and second fastest Brownian
particle in a dendritic spine geometry. The pdf of a particle arriving at the base of the
dendrite within the time $\tau=\tau^{1}+\tau^{2}$ is computed as follows, when the escape time form the head is Poissonian:
\beq
\Pr\{\tau=\tau_{1}+\tau_{2}=t\}=\int\limits_{0}^{t}\Pr\{\tau_2=t-s|\tau_1=s \}\Pr\{\tau_1=s \}ds.
\eeq
The the Markovian property implies that
\beq
\Pr\{\tau_{1}+\tau_{2}=t\}=\int\limits_{0}^{t}\Pr\{\tau_2=t-s\}\Pr\{\tau_1=s\}ds.
\eeq
Using the narrow escape theory \cite{holcmanschuss2015}, the distribution of arrival time of a Brownian particle at the entrance of the dendritic neck is Poissonian,
\beq
\Pr\{ \tau^{1}=s\}  = \gamma e^{-\gamma s},
\eeq
where
 \beqq
\gamma^{-1}=\frac{| \Omega|}{4aD\left[1+\ds\frac{L(\mb{0})+ N(\mb{0})}{2\pi}\,a\log a+o(a\log
a)\right]},
 \eeqq
with $|\Omega|$ the volume of the spherical head, while $a$ is the
radius of the cylindrical neck \cite{holcmanschuss2015} and $L(0)$ and $N(0)$ are the principal mean curvature. After the first particles reaches the cylinder (spine neck), we approximate its Brownian motion in the cylindrical domain  by one-dimensional motion (1D).  By substituting $N=1$ for the first arriving particle, this is given by \eqref{eq:tau1}:
\beq
\Pr\{\tau_{2}=t-s\}=N_R \sum^{\infty}_{n=0} {(-1)^{n}}{\lambda_n}e^{-D \lambda^2_n(t-s)}.
\eeq
Hence,
\beq\label{noreturn}
\Pr\{\tau_{1}+\tau_{2}=t\}&=&\gamma N_R\int\limits_{0}^{t}e^{-\gamma s} \sum^{\infty}_{n=0} {(-1)^{n}}{\lambda_n}e^{-D \lambda^2_n(t-s)}ds\\
&=&\gamma N_R\sum^{\infty}_{n=0}  {(-1)^{n}} \left[  \frac{e^{-D \lambda_n^2 t}-e^{-\gamma t} } {\gamma-D\lambda_n^2} \right].
\eeq
This result represents the pdf of the arrival time of a Brownian particle at the base of a spine, or a process with two time scales: one dictated by diffusion and the other Poissonian.

The result \eqref{noreturn} for the arrival time is derived for the study of the statistics of a single particle. The expression for flux,
\beq
\Phi(t)=\Pr\{\tau_{1}+\tau_{2}=t\},
\eeq
gives
\beq \label{firstspine}
f_{min}(t)=\Pr\{\tau=\min (t_1,\ldots,t_N)=t\}  &=& N \left(1-\int\limits_{0}^{t}\Phi(s)ds\right)^{N-1}\Phi(t)\\
 &=& N\left[J(t) \right]^{N-1}\Phi(t),
\eeq
and using \eqref{noreturn}, we get
\beq \label{Jsurvival}
J(t) =1-\int\limits_{0}^{t}\Phi(s)ds= 1- \gamma \sum^{\infty}_{n=0} \frac{(-1)^n}{(\gamma-D\lambda_n^2)}
 \left[\frac{1-e^{-D \lambda_n^2 t}} {D\lambda_n^2}- \frac{1-e^{-\gamma t}} {\gamma} \right].
\eeq
In the Poissonian approximation, the pdf of the arrival time $\tau^{(2)}$ of the second fastest particle is given by
\beq \label{secondp}
\Pr\{\tau^{(2)}=t\}=N\int\limits_{0}^{t}f_{min}(t-s)f_{min}(s)ds,
\eeq
where $f_{min}(s)=J(s)^{N-1}\Phi(s)$. Thus,
\beq \label{spinetime2}
\Pr\{\tau^{(2)}=t\}=\int\limits_{0}^{t} [J(t-s)]^{N}\Phi(t-s) [J(s)]^{N}\Phi(s)ds.
\eeq
Expressions \eqref{firstspine} and \eqref{spinetime2} represent the distributions of arrival
times of the first and second Brownian particles, initially injected in the spine head
and escape at the end of the cylindrical neck . These expressions are computed under the assumption of no return: when a particle enters the neck cylinder, it cannot return to the head.
%
\subsection{Escape times of Brownian particles with returns to the head} \label{spinestory}
Consider Brownian particles that escape a dendritic spine (Fig. \ref{fig:spinefig4}) into a dendrite with any number of returns to the head after crossing into the neck.  Recrossing is defined to the stochastic separatrix \cite{schuss2013}, the position of which is not known exactly. This effect is likely to impact the first arrival time. The pdf of no return is given by \eqref{noreturn}. To compute the pdf of the shortest escape time $\tau^a$ with returns, we use Bayes' law for the escape density, conditioned on any number of returns, given by
\beq
\Pr\{\tau^{a}=t \}=\sum_{k=0}^{\infty}  \Pr\{\tau^{a}=t|k \} \Pr\{k\},
\eeq
where $\Pr\{k\}=\frac{1}{2^k}$ is the probability that the particle returns $k$ times to the head. The particle hits the stochastic separatrix \cite{DSP} and then returns to the head, before reaching the dendrite. The probability of the escape, conditioned on $k$ returns, $\Pr\{\tau^{a}=t|k \}$, can be computed from the successive arrivals times to the stochastic separatrix, $\tau_1,..\tau_k$, so that
\beq
\Pr\{\tau^{a}=t|k \}=\Pr\{\tau_1+..+\tau_k=t\}.
\eeq
Assuming that the arrival time to the stochastic separatrix is Poissonian with rate $\lambda_S$ \cite{holcmanschuss2015}, we obtain that
\beq
\Pr\{\tau_1+..+\tau_k=t\}=\lambda_S \int\limits_{0}^{t}\frac{(\lambda_Ss)^{n-1}}{(n-1)!} f(t-s)\,ds,
\eeq
where $f(t)$ is the pdf of no return \eqref{noreturn}. Therefore
\beq
\Pr\{\tau^{a}=t \}=\frac{1}{2} f(t)+ \sum_{n=1}^{\infty}  \int\limits_{0}^{t}\lambda_S\frac{(\lambda_Ss)^{n-1}}{(n-1)!} f(t-s) ds\frac{1}{2^k}.
\eeq
Finally,
\beq
\Pr\{\tau^{a}=t \}=\frac{1}{2} f(t)+  \int\limits_{0}^{t}\exp(-\lambda_Ss/2) f(t-s) ds,
\eeq
Expression \eqref{noreturn} with $\lambda_S=\gamma$  gives the final expression for the pdf of the escape time
\beq \label{full}
f_{return}(t)&=&\Pr\{\tau^{a}=t \}\\
&=&\frac{1}{2} f(t)+ \gamma N_R\sum^{\infty}_{n=0}  {(-1)^{n}}\frac{\lambda_n \gamma^2}{4(\gamma-D\lambda_n^2)}
 \left[\frac{e^{-\gamma t/2}-e^{-\gamma t} } {\gamma/2}-  \frac{e^{-\gamma/2 t}-e^{-D \lambda_n^2 t} } {D\lambda_n^2-\gamma} \right].\nonumber
\eeq
The maximum of $f_{return}$ is achieved at the point $t_{max}\approx \frac{2}{\gamma}\log 2$. The pdfs of the
first and second arrivals are computed as
\beqq
f^{(1)}_{min}(t)&=&\Pr\{\tau=min (t_1,\ldots,t_N)=t\} \\
 &=& N \left(1-\int\limits_{0}^{t}f_{return}(s)ds\right)^{N-1}f_{return}(t),
\eeqq
and as in \eqref{secondp},
\beq \label{secondpp}
f^{(2)}_{min}(t)= \Pr\{\tau^{(2)}=t\}=N\int\limits_{0}^{t}f^{(1)}_{min}(t-s)f^{(1)}_{min}(s)ds,
\eeq
The pdfs of the fastest and second fastest arrival times are computed from \eqref{full}, as in the previous
sections. Fig. \ref{fig:spinefig4} shows the pdf of the arrival time $\tau^{(2)}$.  Note that a correction is
needed in \eqref{spinetime2}, because the second particle is not necessarily located inside the head when the
first one arrives at the base.
\begin{figure}[http!]
\centering
\includegraphics[width=.8\textwidth]{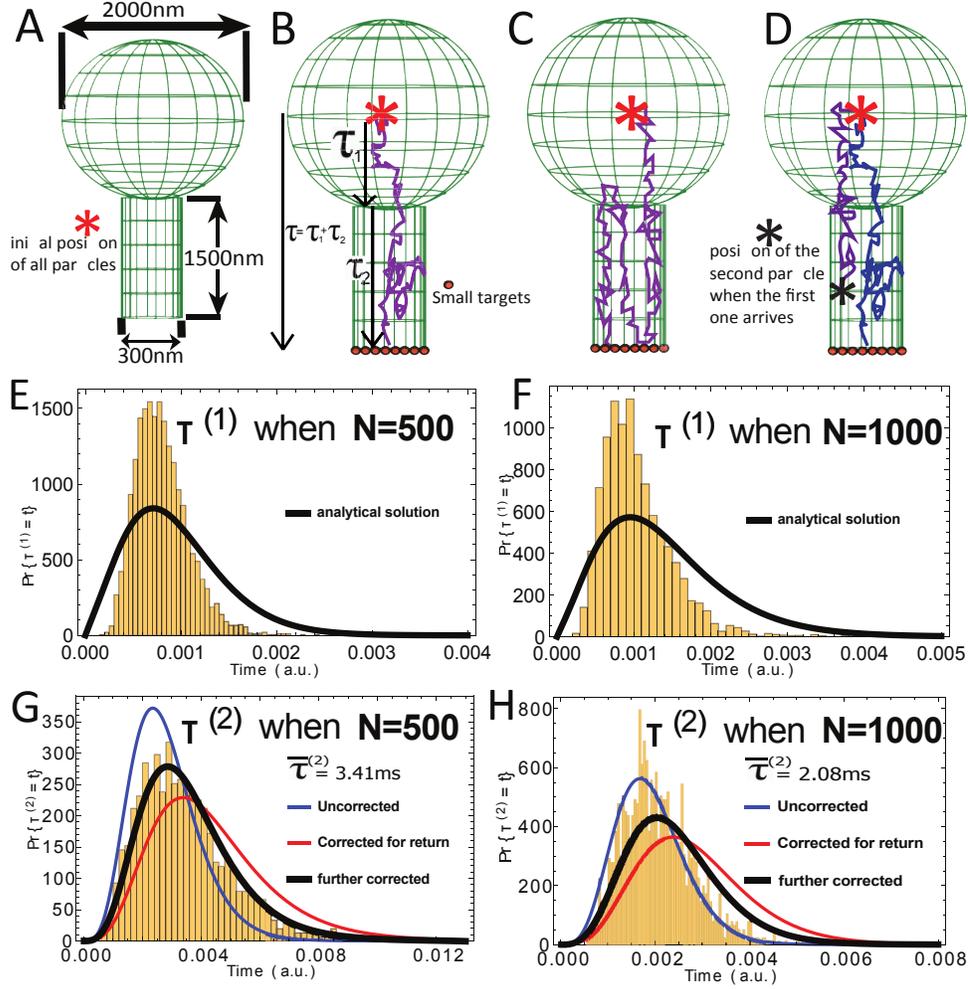}
\caption{\textbf{A-B-C-D:} \small {The geometry of the spine is a spherical head and cylindrical neck. }
Particles are released at the center of the head  and must first reach the top of the neck
and then diffuse through the neck to reach the base. Time taken for each process is represented
with the notation $\tau_1$ and $\tau_2$, making the total time to be $\tau=\tau_1+\tau_2$. C shows a trajectory
that can return to the head. Note that in D, when the first particle arrives, the second one, which could
have return to the head, is now located inside the neck. \textbf{E-F.}  Plot of $\Pr \{\tau^{(1)}=t\}$ for values $N=500, 1000$.
 \textbf{G-H.}  Plot of $\Pr \{\tau^{(2)}=t\}$ for values $N=500, 1000$. {The analytical solution is that of
  \eqref{spinetime2}. Returns are accounted for with \eqref{secondpp}. The further corrected curve (black) is
  given by \ref{Further}. The diffusion coefficient is $D=600 \mu m^2/s$}}
\label{fig:spinefig4}
\end{figure}
Finally, the arrival time formula \ref{secondpp} can be further correct by adding the distribution of the particle inside the head. The correction is similar to the one we obtain in dimension one (formula \ref{approx2} and \ref{approx22}). It can be written
here using the survival probability \ref{Jsurvival} as
\beq \label{Further}
f^{(2)}_{further}(t)= \Pr\{\tau^{(2)}=t\}=N\int\limits_{0}^{t}f^{(1)}_{min}(t-s)f^{(1)}_{min}(s)
J(t)^{N-1}ds.
\eeq
\section{Conclusion and applications of extreme statistics to fast time scale activation in cell biology}
We derived here new asymptotics for the expected arrival time of fastest Brownian particles in several geometries: half a line, a segment, a bounded domain in dimension two and three that contains a small window, and spine-shaped geometry (a ball connected to thin cylinder). We found that the geometry is involved and explored by the fastest particle and the pdf is defined by the shortest ray (with reflections if the is an obstacle \cite{Spivak}) from the source to the target, in contrast with the narrow escape problem \cite{holcmanschuss2015}, where the main geometrical feature is the size of the window and the surface or volume of the domain.  We derived here new laws for the first  arrival time of Brownian
 particles to a target, which can be summarized as
 \beq \label{finalform}
\bar\tau^{d1} &=& \ds\frac{\delta^2_{min}}{4D\ln(\frac{N}{\sqrt{\pi}})}, \hbox{ in  dim } 1\\
\bar{\tau}^{d2} &\approx& \ds \frac{\ds s_2^2}{\ds 4 D \log\left(\frac{\pi \sqrt{2}N}{8\log\left(\frac{1}{\eps}\right)}\right)}, \hbox{ in  dim } 2\\
\bar{\tau}^{3d}&\approx& \frac{\delta^2_m}{2D\sqrt{\log\left(\ds N\frac{4a^2}{\pi^{1/2}\delta^2}\right)}}, \hbox{ in dim } 3,
\eeq
where $\delta_{min}$ is the shortest ray from the source to the window, $D$ is the diffusion coefficient and $N$ is the number of particles, $s_2=|\x-\A|$ and $\x$ is the position of injection and the center of the window is $\A$. Formula \eqref{finalform} is very different from the classical NET, which involves volume or surface area and mean curvature (in dimension 3). When the window is located at the end of a cusp, the asymptotics for the fastest particle are yet unresolved.

We further found that the rate of arrival cannot be approximated as Poissonian, because the fastest particle can arrive at a time scale that can fall into the short-time asymptotic. We further studied here the arrival of a second particle. The mean arrival time of the second can be influence  by the distribution of all particles, especially on a segment, because the distribution of  particles at the time the first particle's arrival is not Dirac's delta function (see section  \ref{s:stat}). However, the case of a spine geometry is interesting, because the first particle may have already arrived, but all other particles can still be in the head (due to the narrow   opening at the neck-head connection).  We further computed the pdf of the arrival time when particle can return to the head after sojourn in the neck (see section \ref{spinestory}).

The present asymptotics  have several important applications: activation of molecular processes are often triggered by the arrival of the first particles (ions or molecules) to target-binding sites.  The simplest model of the motion of calcium ions in cell biology, such as neurons or a dendritic spine (neglecting electrostatic interactions)  is that of independent Brownian particles in a bounded domain. The first two calcium ions that arrive at channels (such as TRP) can trigger the first step of biochemical amplification leading to the photoresponse in fly photoreceptor.
Another example is the activation of a Ryanodine receptor (RyaR),  mediated by the arrival of two calcium ions to the receptor binding sites, which form small targets. Ryanodin receptors are located at the base of the dendritic spine. Computing the distribution of arrival times of Brownian particles at the base, when they are released at the center of the spine head, is a model of calcium release during synaptic activation. Computing the distribution of arrival time reveals that the fastest ions can generate a fast calcium response following synaptic activity. Thus the fastest two calcium ions can cross a sub-cellular structure, thus setting the time scale of activation, which can be much
shorter than the time defined by the classical forward rate, usually computed as the steady-state Brownian flux into the target, or by the narrow escape time \cite{holcmanschuss2015}.


\section*{Acknowledgements}
 This research was supported by the Fondation pour la Recherche M\'edicale - \'Equipes FRM 2016 grant DEQ20160334882.

\bibliography{bibli}
\bibliographystyle{elsarticle-num}

\end{document}